\documentclass{article}

\usepackage{arxiv}

\usepackage[utf8]{inputenc} 
\usepackage[T1]{fontenc}    
\usepackage{hyperref}   
\usepackage{url}           
\usepackage{booktabs}     
\usepackage{amsfonts}       
\usepackage{nicefrac}       
\usepackage{microtype}      
\usepackage{lipsum}
\usepackage{graphicx}
\graphicspath{ {./images/} }
\usepackage{booktabs}
\usepackage{algorithm}
\usepackage{algpseudocode}
\usepackage{wrapfig}
\usepackage{graphicx}
\usepackage{textcomp}
\usepackage{xcolor}
\usepackage{array}
\usepackage{amsthm}
\usepackage{amsmath}
\theoremstyle{definition}
\newtheorem{definition}{Definition}[section]
\newcommand{\framework}{\textbf{GSynPEF}}
\newcommand{\plrv}{PLRV}

\title{\framework{}: A Privacy  Evaluation Framework for Synthetic Graphs}

\author{
 Yu Duan \\
  University of California, Irvine\\
  United States \\
  \texttt{duany12@uci.edu} \\
   \And
 Luca Baldesi\thanks{Work done while at UCI.} \\
  Apple Inc.\\
  United States \\
  \texttt{luca@baldesi.ovh} \\
  \And
 Carter T. Butts \\
  University of California, Irvine\\
  United States \\
  \texttt{buttsc@uci.edu} \\
  \And
 Athina Markopoulou \\
  University of California, Irvine\\
  United States \\
  \texttt{athina@uci.edu} \\
}

\begin{document}
\maketitle

\begin{abstract}
It is often desirable to generate and release synthetic graphs, which resemble real-world data, in a privacy-preserving way.
While substantial progress has been made in designing private graph synthesizers (PGS), opportunities remain for more explicit connections between synthesizer models and privacy evaluation, and better accounting for output variation.
This paper proposes \framework{}, a practical framework for empirical evaluation of privacy and utility in settings for which a real-world graph is given as input and output synthetic graph(s) are observed. 
Our framework links a broad class of graph synthesizers (both statistical and deep-learning based) with privacy metrics, expanding evaluation dimensions to practical utility and risk, through empirical analysis of multiple actual outputs. 
The framework accommodates both an Exact Solution scenario, where synthesizer behavior is directly observable, and a Monte Carlo Approximation scenario where only synthetic outputs from a black box synthesizer are available. 
Extensive experiments on four real-world datasets and seven widely used graph synthesizers demonstrate framework's versatility, enabling privacy evaluation for different attack goals (e.g., to infer an edge or node) and capabilities (e.g., observing one or multiple synthetic graphs).
We analyze evaluation results by examining the relationship between graph statistics and synthesizer design choices, providing insights for data curators in PGS selection and researchers for improved PGS design. 
\end{abstract}


\section{Introduction}
In this paper, we are interested in {\em synthetic graph data} generated using a model or algorithm (which we refer to as a {\em graph synthesizer}), intended to resemble observed data from real-world systems. 
While there are many kinds of synthetic data, we focus here on \emph{\em proxy observations,} in which some observed dataset $Y$ is used to generate a synthetic dataset $Y'$ that preserves as many features of $Y$ as possible while altering potentially sensitive details. 
Such datasets are widely used in different fields (e.g., healthcare~\cite{gonzales2023synthetic}, finance~\cite{assefa2020generating}) to address challenges including data scarcity, privacy concerns~\cite{el2020practical}, and data bias. 
Data curators often face a dilemma: while regulations like GDPR, CCPA, and HIPAA mandate strict protection of individual privacy and limit data sharing, data scientists/engineers require large volumes of high-quality data to effectively develop/evaluate ML models. Further (and somewhat paradoxically), researchers face increasing demands from funders and publishers to publicly release data, even as the above privacy requirements continue to tighten.
Synthetic data provide a possible solution to this dilemma, by enabling the creation of proxy observations that preserve important statistical properties of real data while mitigating privacy risks, thereby supporting both regulatory compliance and data-driven research/applications.  

\textbf{Synthetic Graph Use Cases.} 
Among different data types, graph data are particularly valuable, as they represent relationships between entities and embed rich structural information.
In social networks, synthetic graphs are used to model user interactions or information diffusion~\cite{jin2018information}, which allow researchers to test new recommendation systems or study social phenomena without using real social graphs.
In network communication~\cite{bakhshaliyev2020generation,xu2024anonymization}, synthetic graphs represent network topology, aiding the development and testing of routing algorithms and network resilience strategies. 
In finance, synthetic transaction graphs~\cite{assefa2020generating} train fraud detection model and simulate user behavior without leaking real transactions or customer IDs.
Healthcare synthetic data can represent doctor patient relationships~\cite{paudel2017detection}, which can be used to detect anomaly/fraud.

\textbf{Privacy Risks in Graph Data.} 
It is well-known that privacy risks can exist in graph datasets even after explicit identifiers are removed. 
For instance, \cite{backstrom2007wherefore} describes an attack family to infer whether edges exist between two target nodes from graph structure.
The algorithm in \cite{narayanan2009anonymizing} can successfully re-identify users in anonymous Twitter graphs with auxiliary information. 
\cite{chen2020active} points out that re-identification attacks exist in identifier-free datasets, and risk is amplified in a dynamic graph setting. 
Various methods have been proposed to mitigate such privacy risks from graph topology.
\cite{sala2011sharing} introduces a model to publish synthetic graphs that preserve degree correlation. 
\cite{casas2013algorithm} proposes a K-degree graph model to anonymize graphs by structure modification.
\cite{qin2017generating,yuan2023privgraph} develop PGS to generate synthetic graphs that preserve community structures, under differential privacy (DP).  
\cite{liu2024pgb} summarizes current PGS in terms of three core components: representation, perturbation and construction.
However, privacy risks remain in synthetic graph outputs, as rich structural information enables adversaries to infer relationships or re-identify individuals. 
\cite{wu2024provable} shows that synthetic graphs can leak sensitive information due to structural vulnerabilities, highlighting privacy risks that remain even when privacy defenses are applied (e.g., adding noise).  
\cite{yuan2024unveiling} shows that graph structures play a key role in privacy leakage by proposing a new attack model, allowing inference for private attributes.

\textbf{Our Goal.}
Compared to previous work~\cite{sala2011sharing,casas2013algorithm,qin2017generating,yuan2023privgraph,liu2024pgb}, whose goal is to design better PGS algorithms (e.g., using DP), our goal is to build a practical framework to evaluate the privacy risk when using graph synthesizers.

Currently, there is no practical framework to evaluate privacy and utility for graph synthesizers intended to produce proxy observations of graphs, an important application setting for data stakeholders.
The relation between graph synthesizers and their privacy guarantees is not well understood by practitioners, making it unclear which models deliver better privacy foundations. 
Such privacy evaluation may be needed by different users, including data contributors (those who contribute the real-world data), data curators, or researchers (who use the synthesizers and release the outputs). 
Depending on the use case, we may be interested in evaluating the privacy risk for a particular or all inputs (whole graph), one or all possible outputs, of the graph synthesizer.
We may also consider different attack goals, including inferring an edge or node of the original graph, based on the available observations.
Moreover, since randomness is inherent in the graph synthesis process, evaluations should be conducted based on the actual generated output.
The graph data type further increases the computational complexity of  conducting the evaluation.

\begin{figure}[t] 
  \centering
  \includegraphics[width=0.5\linewidth, alt = {Conceptual diagram illustrating a framework pipeline. An original graph G is fed into a graph synthesizer model M, which outputs a synthetic proxy graph o. They are evaluated across privacy and utility metrics.}]{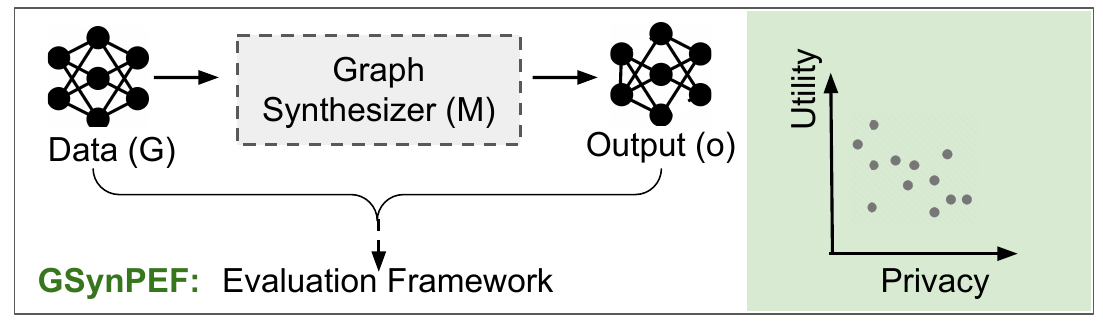}
  \caption{\textbf{Conceptual Diagram.} Input of graph synthesizer ($M$) is original graph ($G$), and output is synthetic graph ($o$) intended as a proxy generated for $G$. Proposed framework is able to evaluate a setup ($G \times M$) from privacy/utility, with actual outputs of $M(G)$.} 
  \label{fig:intro-mini}
\end{figure}

Our framework aims to address these challenges in evaluation of graph synthesizers for proxy observations. 
A conceptual diagram is shown in Figure~\ref{fig:intro-mini}, highlighting the evaluation object (a setup, composed of dataset $G$ and mechanism $M$), and key dimensions in our evaluation. 

In summary, we make the following contributions.

First, we propose \framework{}, a framework for evaluating graph synthesizers on the  key dimensions of privacy/utility, and we interpret privacy utility trade-off as risk. We leverage the privacy loss random variable and approximate differential privacy to bridge the gap between graph synthesizer and privacy evaluation.

Second, we offer two evaluation scenarios (Exact Solution and Monte Carlo Approximation) to simulate different model access capabilities. Using actual outputs, we compute privacy loss random variables in the neighbor-by-output space. Then we extract  metrics at different granularities, which align with different attacks goals (edge/node) and attack capabilities (observing one or multiple synthetic graphs), serving for different use cases.

Third, we implemented \framework{} across setups, then scaled up experiments to large graph, with Monte Carlo sampling estimation and uncertainty quantification. In total, we evaluated seven different graphs synthesizers that span in two different model families, statistical and deep learning based, with four real-world datasets. We observed that synthesizer performance vary across  datasets, reflecting an interaction between graph statistics and synthesizers' design. Lastly, as data size increases, utility error metrics become more pronounced, and the privacy evaluation ranking may shift.
 
The rest of this paper is organized as follows. Firstly we list the preliminaries related to our framework in Section~\ref{sec:preliminaries}. Next we present major phases of \framework{} framework in Section~\ref{sec:framework}. Experiments results and findings are shown in Section~\ref{sec:experiment-and-evalustion}.
At last, we summarize the related work in Section~\ref{sec:related-work}, discuss limitations or guidelines in Section~\ref{sec:discussion/limitation}, conclude this work in Section~\ref{sec:conclusion}.

\section{Preliminaries \& Problem Statement}
\label{sec:preliminaries}
\subsection{Preliminaries}

For an undirected, unweighted graph $G = (V, E)$, we have the following definitions related to privacy measurement.

\begin{definition}[Graph Edge Neighbor]
If two graphs $G, G'$ are only different in one edge, they are called graph edge neighbors~\cite{li2023private, yuan2023privgraph}. 
The symmetric difference $x \oplus y$ is the sets of elements in either set $x$ or set $y$, but not in both, i.e., $x \oplus y = (x \cup y) \backslash (x \cap y)$.
Two graphs $G = \{V, E\}$, $G' = \{V', E'\}$ are neighbors if they satisfy Equation~\ref{equ:e-neighbor}. 
In other words, one graph can be obtained from the other by either adding or removing one edge. 

\begin{equation}
\label{equ:e-neighbor}
    |V \oplus V'|  +  |E \oplus E'| = 1
\end{equation}
\end{definition}

\begin{definition}[Graph Differential Privacy]
For a graph neighbor pair $(G, G')$, consider a randomized graph generating mechanism $M$ with possible outcomes $Range(M)$. 
If $M$ satisfies Equation~\ref{equ:gdp}, we say that $M$ is $\epsilon$-differentially private for graph $G$~\cite{dwork2006differential, wang2013preserving}.  
Intuitively, $M$ preserves the privacy of $G$ to the extent that there is no output of $M(G)$ that is substantially more indicative of $G$ than of its neighbors.
\begin{equation}
\label{equ:gdp}
    \forall T \subseteq Range(M): Pr[M(G) \in T] \leq e^{\epsilon} \times Pr[M(G') \in T]
\end{equation}
\end{definition}

\begin{definition}[Privacy Loss Random Variable]
\label{def:plrv}
Privacy loss random variable (\plrv{}) is proposed in \cite{bun2016concentrated}. It is defined by the logarithm of probability ratio between two neighbor datasets to produce a specific output $o$ under mechanism $M$ (definition in Equation~\ref{equ:plrv}). 
$M$ is $\epsilon$-differential private if and only if $Pr[$\plrv{}$>\epsilon]=0$.
Additionally, $M$ being $(\epsilon, \delta)$-differentially private is equivalent to the requirement that $Pr[$\plrv{}$> \epsilon] \leq \delta$. This is also called \emph{approximate differential privacy}~\cite{asoodeh2021three, dwork2006calibrating}, which is a relaxation of $\epsilon$-DP. 
We apply it on graph data.

\begin{equation}
\label{equ:plrv}
    I(o, G') 
    =  
    \ln{\frac{Pr[M(G) = o]}{Pr[M(G') = o]}}
\end{equation}
\end{definition}

\textbf{\plrv{} vs. $\epsilon$}
\plrv{} measures the empirical privacy loss for a specific output and data neighbor pair, random variable reflects the mechanism ($M$) randomness over possible outputs. $\epsilon$ is a global upper bound on \plrv{} values. 
$\epsilon$ guarantees that no individual privacy loss exceeds this bound in pure differential privacy (or rarely, with probability $\delta$, in approximate differential privacy). 
In our work, we compute \plrv{} values following Definition~\ref{def:plrv} for different graph neighbor pairs and observed outputs. 
Next, we apply $({\epsilon}, {\delta})$-differential privacy to the computed values, with a fixed small ${\delta}$. 
We then determine the corresponding ${\epsilon}$ satisfying this privacy guarantee, referred to here as the $\delta$-bound. 

\subsection{Problem Description \& Solution}
We are interested in a graph synthesizer $M$, which maps input graph $G$ to proxy observation $o$ with randomness. Utility for $o$ is determined by the extent to which it preserves features of $G$, while privacy is determined by the extent to which sensitive details of $G$ are obfuscated. Assessment of alternative choices of $M$ depends on having a systematic way to evaluate this privacy/utility trade-off. At present, such an evaluation scheme is not available for graph data. This gap leaves data curators and researchers without reliable tools for mechanism selection or for guiding the design of effective PGS methods.  
We are particularly interested in understanding: \emph{how much privacy loss and utility error arise from a specific setup ($G\times M$) ? Moreover, how robust are these evaluation results across outputs?} 
Our work is designed to systematically address these questions by proposing an evaluation framework based on actual outputs.

We propose \framework{} to answer these questions. 
This framework uses \plrv{} for privacy loss and leverages the $\delta$-bound in approximate DP to extract key metrics. 
The framework supports privacy evaluation from different granularities and two practical scenarios. 
It also integrates utility and risk analysis to provide a comprehensive assessment.
To improve scalability in large graph computations, we incorporate sampling to generate privacy loss and utility error values then extract metrics. 
This framework enables data curators to select appropriate mechanisms and guide researchers in designing advanced PGS methods.

\section{Framework: \framework{}}
\label{sec:framework}

\begin{figure*}[!t] 
  \centering
  \includegraphics[width=\linewidth, alt = {High-level overview diagram of the framework pipeline divided into three main phases: Generation, Computation, and Evaluation. In the Generation phase, an input dataset G and synthesizer model M produce graph neighbors and multiple outputs across Edge Sampled and Monte Carlo Approximation scenarios. In the Computation phase, privacy loss values are mapped into an image space grid alongside label-dependent and label-independent utility errors. In the Evaluation phase, metrics are extracted across node or edge privacy goals and per-output or per-setup attack capabilities.}]{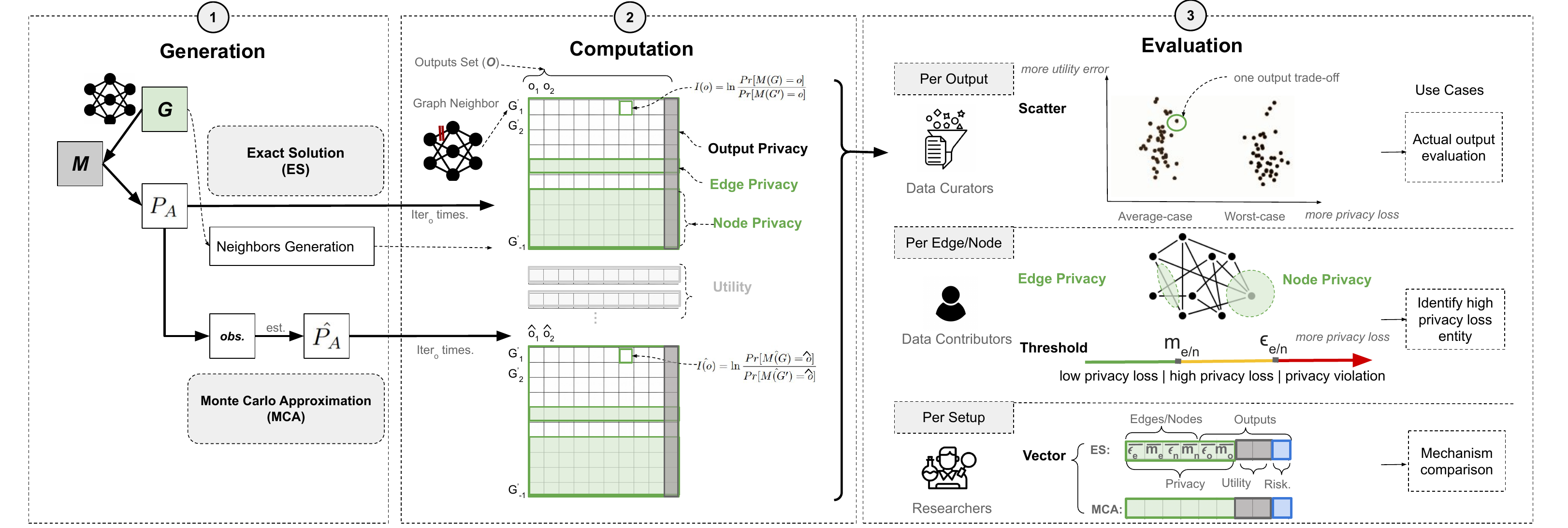}
  \caption{\textbf{Overview of \framework{}.} Input is a ``setup'', which includes a dataset $G$ and a graph synthesizer $M$. Outputs are evaluation results, from different granularity levels by considering privacy/utility dimensions. \framework{} incorporates three phases: generation, computation and evaluation. 
  (i) The generation phase involves two parallel steps: generation of graph neighbors and multiple outputs for a given setup. 
  ES scenario models privacy loss to an adversary who has access to edge probability matrix $P_A$, while MCA scenario models privacy loss to an adversary who only gets to see some number of draws from $M$ with the target dataset (Section~\ref{sec:generation}).
  (ii) The computation phase describes the process to compute privacy loss and utility error values. 
  Privacy loss is computed as \plrv{}s over the computation space $\mathbb{Z}$, where each cell corresponds to an output–neighbor pair and collectively forms matrix $I_{\plrv{}}$. 
  Utility error is evaluated using both label-dependent and label-independent metrics (Section~\ref{sec:computation}).
  (iii) The evaluation phase extracts metrics from computed values with various granularities. 
  Edge/node privacy aligns with different attack goals, and per output/setup privacy corresponds to different attack capabilities (Section~\ref{sec:evaluation}).}
  \label{fig:overview}
\end{figure*}

\subsection{Overview}
\label{sec:method:overview}
Figure~\ref{fig:overview} illustrates the \framework{} framework, which includes three phases: generation, computation and evaluation. Input is a setup, which consists of dataset $G$ and mechanism $M$, and output consists of evaluation results. 
First, in two scenarios (Exact Solution and Monte Carlo Approximation), we generate outputs ($O$) and graph neighbors ($G'$) to build the computational space ($\mathbb{Z}$). 
Second, we compute PLRV matrix $I_{\plrv{}}$ for privacy loss values,
and compute utility error values for each output, including both label independent and label dependent values.
Third, we transform computed values into key metrics with different granularity levels, per output, edge/node and setup.
Finally, we scale up the method to large graph.
We summarize all the used notations in Table~\ref{tab:notation}.

\begin{table*}[!t] 
    \caption{Notations}
    \begin{center}
    \begin{tabular}
    {p{1.5cm}|p{12cm}}
        \toprule
        \textbf{Notation} & \textbf{Description} \\ 
        \midrule
        $M$         & A randomized mechanism (a graph synthesizer with randomness)  \\
        $G(V,E)$              & Original graph with nodes set $V$ and edge set $E$.\\
        $G'$              & Edge neighbor of original graph \\
        $O$              & Output set of $M(G)$, with length $Iter_o$. $O$ is $[o_1, o_2,... o_{-1}]$ \\ 
        $A$         & Adjacency matrix of original graph\\ 
        $N$         & Node number in $G$ \\  
        $P_A$ & The edge probability matrix that generates the synthetic graph, dimension is $N\times N$.\\
        $\hat{P_A}$ & The edge probability matrix that is estimated by several observations of $M(G)$.\\
        \midrule
        $\mathbb{Z}$
        & Computation space, which dimension is [$\frac{N(N-1)}{2}, Iter_o$].\\
        $I(G', o)$ 
        & Privacy loss random variable computed for one edge neighbor $G'$ with one observed output $o$.\\
        $I_{PLRV}$ 
        & \plrv{} matrix, dimension is the number of $G'$ $\times$ $Iter_o$, each cell is $I(G', o)$.\\
        $(pl_i, ue_i)$         & Privacy loss and utility error for one output $o_i$. \\
        $\delta$ 
        & Privacy loss breach probability for worst-case.\\
        ${\epsilon}_o$         & Worst-case privacy loss for one output $o$. \\
        $m_o$         & Average-case privacy loss for one output $o$. \\ 
        $\overline{m_o}$         & Average privacy loss in average case across all the outputs in output set $O$. \\
        $\hat{\overline{m_o}}$         & Estimated average privacy loss in average case across all the outputs in output set $O$. \\
        \midrule
        $B$         & Modularity matrix. \\
        \textbf{V}         & List of the eigenvectors.\\
        $\Lambda$         & List of the eigenvalues.\\
        $\alpha$         & Percent of the eigenvalues to preserve. \\
        $\beta$ & Bernoulli process, $P_A \rightarrow o$ \\
        \bottomrule
    \end{tabular}
    \end{center}
    \label{tab:notation}
\end{table*}

\subsection{Generation Phase}
\label{sec:generation}

\begin{algorithm}[!t] 
    \caption{Graph Synthesizer $M$ (e.g., SGF)}
    \label{alg:sgf}
    \begin{algorithmic}[1] 
        \State \textbf{Input:} $G$ 
        \State \textbf{Output:} $P_A$ \Comment{\small{$P_A$ is the edge probability matrix of A}}
        \State $A = Adj(G)$ \Comment{\small{Adjacency matrix} }
        \State $n = len(A)$
        \State $B = A - \frac{K K^T}{\sum K}$ \Comment{Modularity matrix}
        \State $\Lambda, \bf{V} = \textbf{decompose}(B)$ \Comment{\small{Eigenvalues/eigenvectors}}
        \State $idx \gets \text{Sorted}(idx, |\Lambda|)$ \Comment{\small{Sort $idx$ by $|\Lambda|$ in desce. order}}
        \State $\bf{\Lambda}[\alpha n: n] = 0$ \Comment{\small{$\alpha$ is the inherent param. of $M$}}
        \State $\widetilde{B} = \bf{V}{\Lambda}\bf{V^T} $
        \State $\widetilde{A} = \widetilde{B} + \frac{K K^T}{\sum K} $
        \State $P_A = Clip(\widetilde{A})$ with [0,1]
        \State \textbf{return} $P_A$
    \end{algorithmic}
\end{algorithm}

We consider two scenarios in the generation phase, which simulate adversaries with different access capabilities to $M$. 
Each scenario involves two independent steps, generations of graph neighbors ($G'$) and output ($o$) set.

\textbf{Exact Solution (ES).}
This scenario models an adversary who has access to the edge probability matrix $P_A$ used by $M$ to produce proxy observations, enabling direct assessment of the underlying generative process.   
First, we generate all the edge neighbors of original graph. Second, we acquire $P_A$ and run Bernoulli Process $Iter_o$ times to generate multiple outputs. 
After generation, we build a neighbor-by-output computational space $\mathbb{Z}$ as blueprint, to compute a \plrv{} in each cell.

\textbf{Monte Carlo Approximation (MCA).}
In a more restrictive scenario, the adversary lacks access to the edge probability matrix $P_A$ and only can observe a limited number of synthetic graph outputs (possibly only one, in a typical proxy data setting). 
This limitation forces the evaluation to rely solely on observed outputs, reflecting a more realistic scenario where $P_A$ is neither available nor disclosed.
We refer to this scenario as Monte Carlo Approximation (MCA), which estimates edge probability matrix as $\hat{P}_A$ by averaging edge occurrences across multiple observations.

The major difference between two scenarios is that ES has access to $P_A$ used by $M$, while MCA can only observe multiple outputs of $M(G)$ to estimate the underlying edge probability matrix $\hat{P_A}$. 
ES is thus a scenario can provide more precise evaluation, while MCA is a more practical scenario which models the access to actual outputs. 
In Algorithm~\ref{alg:sgf}, we use spectral graph forge synthesizer (SGF)~\cite{baldesi2018spectral} as an example, to show how to get $P_A$ from $G$ and $M$ in ES scenario. Additionally, in both scenarios, we traverse all the possible graph edge neighbors to build the computational space for \plrv{}.
Our evaluation framework makes the following assumptions for the graph synthesizer $M$: (i) both input and output are graphs, which is a practical setting for graph data release; (ii) the mechanism functions as a Bernoulli graph synthesizer, generating edges independently, which is widely used in PGS design shown in~\cite{liu2024pgb}; (iii) node correspondence between original graph and synthetic output is given. 
If the latter does not exist, we relax this assumption by employing only label-independent metrics for utility evaluation.

\subsection{Computation Phase}
\label{sec:computation}
We show how to compute the privacy loss and utility error values, then scale up the current computation to large graph.

\subsubsection{Computing Values}
\label{sec:computation-metric}

\textbf{Privacy Loss.}
For privacy, we can compute \plrv{} values for two different scenarios.
In the ES scenario, we have access to $P_A$, which is used to generate $Iter_o$ outputs, then compute \plrv{} in neighbor-by-output space.
We define the neighbor-by-output space as computation space $\mathbb{Z}$, a two-dimensional space spanning graph neighbors and outputs.
Algorithm~\ref{alg:quantifyeneis} shows how we compute \plrv{} matrix ($I_{\plrv{}}$) in ES scenario, computation complexity is $O(N^4Iter_o)$. ES scenario captures synthetic graphs that appear with probability greater than $\frac{1}{Iter_o}$ in output.
Due to limited computation resources in practice, we select a reasonable $Iter_o$ for experiment (see  Section~\ref{sec:hyper-params}).
Similarly, for MCA, the evaluation framework can access multiple outputs to calculate $\hat{P_A}$ (estimated $P_A$), then compute corresponding $I_{\plrv{}}$ for further evaluation.

\textbf{Utility Error.}
For utility error, we consider both label independent and label dependent categories. The error is based on the discrepancy between $f(G)$ and $f(M(G))$, and the error associated with treating $M(G)$ as a proxy for $G$. 
Label dependent utility error metrics provide a more detailed assessment of graph structure based on known node correspondence (e.g. degree sequence), while label independent utility error metrics do not change with node permutation (e.g. degree distribution). 
In the label independent category, we consider three widely-used graph structure descriptions: modularity, clustering coefficient, and degree distribution. 
For modularity and clustering coefficient, we use their relative error (values range between 0 and 1); for degree distribution, we use KL divergence (values are non-negative and can exceed 1). 
These three utility error metrics are denoted as $ue_{LI} = [ue_1, ue_2, ue_3]$ for each output.
In the label dependent category, we choose degree sequence, eigenvector centrality, and closeness centrality. 
For each metric, we compute the correlation between the utility vectors of the original and synthetic graphs, and define utility error as $(1-cor) \in [0,2]$, denoted as $ue_{LD} = [ue_4, ue_5, ue_6]$ for each output.

\subsubsection{Scaling Up to Large Graphs}
\label{sec: method-lg}
Exhaustive PLRV calculation over whole computation space $\mathbb{Z}$ is infeasible due to computation complexity (Section~\ref{sec:computation-metric}). 
So we modified Algorithm~\ref{alg:quantifyeneis} to use sampling-based estimation, computing both the average metric and its uncertainty.

\textbf{Monte Carlo Sampling.}
This is particularly important to enable privacy evaluations of large graphs. To that end,  we modify Algorithm~\ref{alg:quantifyeneis} to use Monte Carlo sampling\footnote{
Please notice that the MC sampling used here, is different from MC used in Section~\ref{sec:generation}. This Section~\ref{sec: method-lg}'s goal is to make our privacy evaluation framework more efficient in large graph dataset, with a modified version of Algorithm~\ref{alg:quantifyeneis}.
In particular, lines 8-16 in Algorithm~\ref{alg:quantifyeneis} perform exhaustive traversal in the computation space $\mathbb{Z}$. 
The modified version of Algorithm~\ref{alg:quantifyeneis} applies Monte Carlo sampling in $\mathbb{Z}$, then computes PLRVs in this subspace $z_n$, which replaces lines 7-17 with line 18-28.}
We draw $n$ samples, $z^{(1)}, z^{(2)}, \dots, z^{(n)}$, uniformly at random from the computation space $\mathbb{Z}$, which encompasses all graph neighbors and observed outputs.
The sum of the samples' quantity can be expressed as $T=\sum_{z \in \mathbb{Z}} f(z) = |\mathbb{Z}|\mathbf{E} f(Z)$, the sample mean $\overline{f(Z_n)} = 1/n \sum_i f(z^{(i)}) \to \mathbf{E} f(Z)$ as $n \to \infty$ (by the law of large numbers).
In our work, $z$ is a randomized position located in computation space $\mathbb{Z}$, and $f(z)$ is the \plrv{} computed for this position.

\textbf{Uncertainty Quantification.}
To measure the uncertainty within estimation, we follow the Type A uncertainty calculation in ~\cite{jcgm:2008:EMDG}, then normalize the standard uncertainty with sample mean $\mu$. The relative uncertainty score is calculated by $\text{RU} = \frac{s}{\mu \sqrt{n}}$ ($s^2$ is the variance and $n$ is the amount of independent observations).

\begin{algorithm}[t!] 
    \caption{Privacy Evaluation, Exact Solution Scenario}
    \label{alg:quantifyeneis}
    \footnotesize
    \begin{algorithmic}[1] 
        \State \textbf{Input:} $G$, $M$ \Comment{\small{A setup}}
        \State \textbf{Output:} $I_{\plrv{}}$ \Comment{\small{\plrv{} matrix}}

        \State $A = Adj(G)$ 
        \State $P_A = M(A)$
        \State $A^{'}_{ls} = graph\rule{0.1cm}{0.5pt}neighbors\rule{0.1cm}{0.5pt}generation\ A$
        \State $O = generate\rule{0.1cm}{0.5pt}random\rule{0.1cm}{0.5pt}samples\rule{0.1cm}{0.5pt}from\ (P_A, Iter_o)$
        \State $I_{\plrv{}} = list()$ \Comment{\small{Dimension = [$\frac{N(N-1)}{2}$, $Iter_o$]}}
        \For{$A'$ in $A'_{ls}$ } \Comment{\small{Outer loop: complexity = $O(N^2)$}}
        \State $I_{ls} = list()$
        \State $P_{A'} = M(A')$
        \For{$o$ in $O$ } \Comment{\small{Inner loop: complexity = $O(Iter_o)$}}
        \State $I(o, G') =  \ln \{\frac{Pr[\beta(P_A) = o]}{Pr[\beta(P_{A'})) = o]} \} $ 
        \State $I_{ls}.append(I(o, G'))$
        \EndFor
        \State $I_{\plrv{}}.append(I_{ls})$
        \EndFor
        \State \textbf{return} $I_{\plrv{}}$
        \Statex \hrulefill
        \Statex \textbf{Modified for Large Graph} (same lines 1-6, substitute lines 7 - 17)
        \State $G^{'}_{ls} = graph\rule{0.1cm}{0.5pt}neighbors\rule{0.1cm}{0.5pt}generation\ G$
        \State Dimension ($\mathbb{Z}$) = [$\frac{N(N-1)}{2}$, $Iter_o$]
        \State $z_n \gets$ Randomly sample $n$ cells from computation space $\mathbb{Z}$
        \State $I_{\plrv{}} = list()$ 
        \For{each cell $(i,j) \in z_n$}
            \State $o = O[j]$
            \State $A'= A'_{ls}[i]$, $G'= G'_{ls}[i]$
            \State $I(o, G') =  \ln \{\frac{Pr[\beta(P_A) = o]}{Pr[\beta(P_{A'})) = o]} \} $ 
            \State $I_{PLRV}.append(I(o, G'))$
        \EndFor
        \State \textbf{return} $I_{\plrv{}}$
    \end{algorithmic}
\end{algorithm}

\subsection{Evaluation Phase}
\label{sec:evaluation}

We now consider to extract key metrics from computed values in privacy/utility, evaluating each output, edge/node and setup.
We show the evaluation metrics/vector details in Section~\ref{sec:method:evaluation metrics}-\ref{sec:method:evaluation vector}, then show the metrics adapted to large graph in Section~\ref{sec:method:evaluation metrics-lg}.

\subsubsection{Evaluation Metrics}
\label{sec:method:evaluation metrics}
\begin{figure}[t!] 
  \centering
  \includegraphics[width=0.6\linewidth, alt={Diagram illustrates the evaluation vector formation pipeline, converting computed privacy and utility values from a raw matrix representation into structured evaluation vectors and a final privacy versus utility trade-off scatter plot.}]{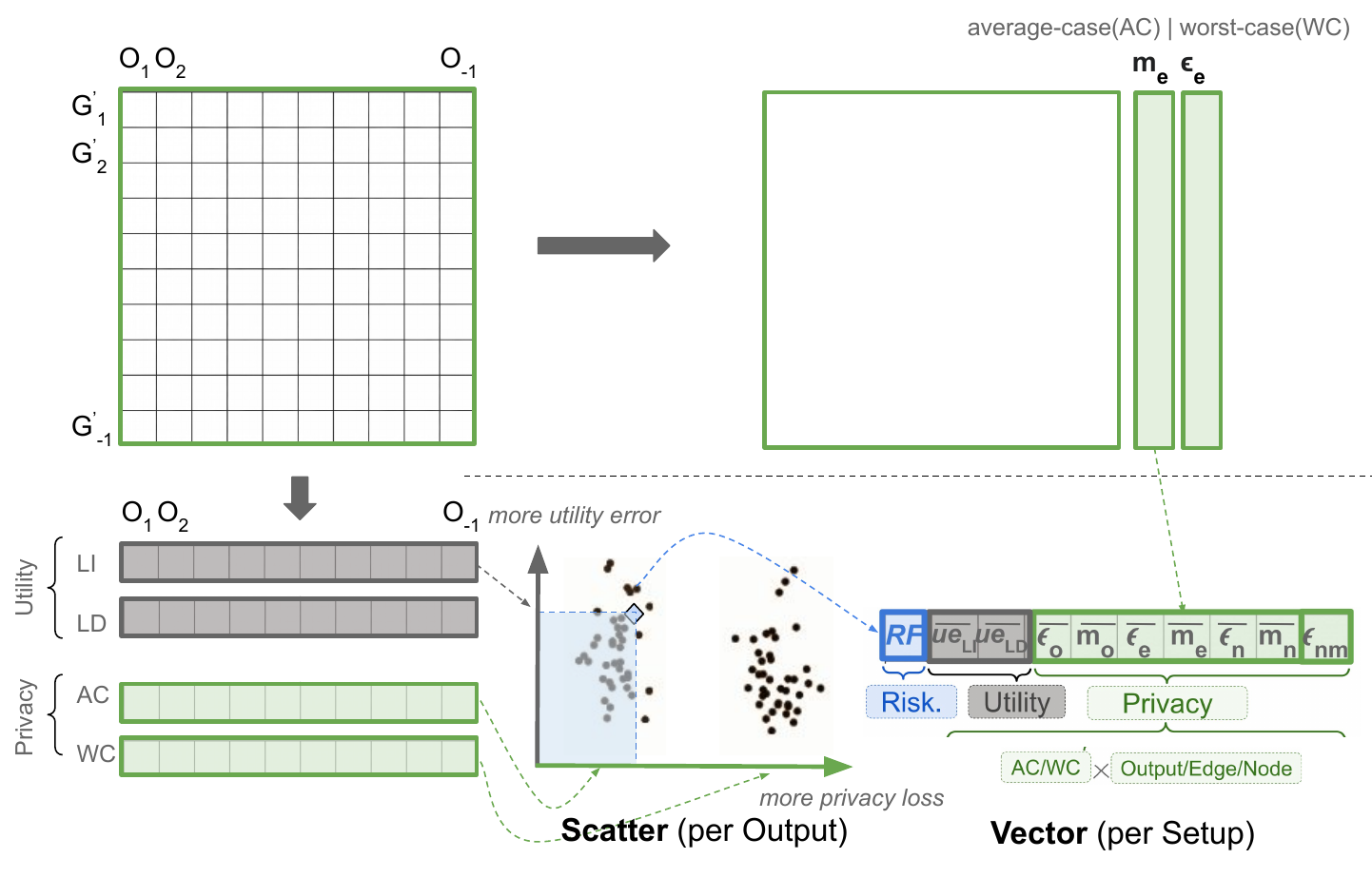}
  \caption{\textbf{Evaluation Vector Formation.} From computed privacy/utility values to evaluation vector and scatter plot.}
  \label{fig:metric-extraction}
\end{figure}

Different combinations in computed values are appropriate for different use cases, e.g., by data curators, data contributors, researchers.
Then we present these results with intuitive visualizations.
Figure~\ref{fig:metric-extraction} illustrates how we extract the evaluation vector from the computed values.
Edge or node privacy aligns with different attack goals (e.g., edge or node), and per output or setup privacy corresponds to different attack capabilities (e.g., observing one or multiple synthetic graphs).

\textbf{Per Output/Edge/Node.}
Evaluation results of each output are visualized in scatter plots. Output utility is evaluated by utility error. Output privacy involves \plrv{}s for one output across graph neighbors.
\plrv{} $I(o,G')$ measures how much an adversary can infer one edge’s true presence/absence from an observed output.
Output privacy quantifies how well an observed output hides the presence/absence of edges in the original graph.
It is measured by aggregating privacy loss for a given output across all edge neighbors, $m_o = \frac{\sum^{|E|}_{j=1}I(o,G'_j)}{|E|}$ for average case and $\epsilon_o$ for worst case ($\delta$-bound) per output.
High output privacy loss means an adversary can more easily infer the presence/absence of edges from a specific output.

Edge privacy considers privacy loss for one edge neighbor across outputs. 
Node privacy considers privacy loss from relevant edge neighbors across outputs. 
Edge privacy loss measures inference risks for a specific edge across $Iter_o$ observed outputs. By observing those outputs, how much certainty an adversary can infer about whether an edge truly exists in the original graph.
High edge privacy loss means that an adversary is more certain about edge existence in original graph, which is a worse privacy level.
Node privacy loss is associated with the edges that are connected to the specific node.
We also consider average case (mean) and worst case ($\delta$-bound) for each entity (edge/node), denoted as $[m_e, \epsilon_e]$ for each edge, $[m_n, \epsilon_n]$ for each node.
Average case values are defined by $m_e = \frac{\sum^{Iter_o}_{i=1}I(o_i,G')}{Iter_o}$ and $ m_n =\frac{\sum^{Iter_o}_{i=1}\sum^{|E_n|}_{j=1}I(o_i,G'_j)}{|E_n|*Iter_o}.$ 

\subsubsection{Evaluation Vector.}
\label{sec:method:evaluation vector}
Recall that input to our framework is a {\em setup} (a dataset $G$ and a mechanism $M$, see Section~\ref{sec:method:overview}).
Each setup is evaluated by a privacy-utility vector, and {\em risk fraction (RF)}. 

\textbf{Privacy-utility Vector.} This vector summarizes evaluation results from all observed outputs, and privacy evaluations from existing edges/nodes. 
In the ES scenario, we build $I_{\plrv{}}$ matrix, each cell is a PLRV for one graph neighbor with one observed synthetic graph. 
We also derive the $\delta$-bound of \plrv{}s in computation space $\mathbb{Z}$ that considers all edge neighbors and all observed outputs, denoted as $\epsilon_{mn}$. This metric enables us to assess the worst-case privacy loss by considering the whole computation space $\mathbb{Z}$. 
To sum up, the privacy loss vector is denoted as $[\overline{\epsilon_o}, \overline{m_o}, \overline{\epsilon_e}, \overline{m_e}, \overline{\epsilon_n}, \overline{m_n}, \epsilon_{mn}]$; the utility error vector is denoted as $[\overline{ue_1}, \overline{ue_2}, \overline{ue_3}, \overline{ue_4}, \overline{ue_5}, \overline{ue_6}]$. Overlines indicate average values across outputs/edges/nodes.

\textbf{Risk Fraction (RF).} We define RF as the proportion of outputs that exceed the specified loss constraints $(pl_c, ue_c)$, and quantifies the risk/likelihood that the privacy-utility trade-off of one observed output surpasses the constraints.
Equation~\ref{equ:rf} defines $RF$ for a setup, the indicator variable $1(pl \leq pl_c, ue \leq ue_c)$ counts the number of outputs satisfying the constraints. Smaller RF implies that a majority of output evaluations satisfy the constraints. $RF_i$ corresponds to $ue_i$ for one setup within observed outputs, $RF$ = $[RF_1, RF_2, RF_3, RF_4, RF_5, RF_6]$ for six utility metrics.
Specifically, we choose fixed constant constraint values for consistent comparison across $M$. 
Then we assess the risk of each output $(pl_i, ue_i)$ by its average-case privacy metric ($m_o$) and utility metric, as average-case privacy is more representative. 
Next we aggregate the risk of all the observed output to evaluate the setup risk.
Additionally, by varying the constant constraints in two dimensions, RF produces a heatmap for each $M$, which facilitates the comparison.

\begin{equation}
\label{equ:rf}
    RF
    = 
    1 - \frac{\sum_{i=1}^{Iter_o}{1(pl_i \leq pl_c, ue_i \leq ue_c)}}{Iter_o}
\end{equation}

\subsubsection{Large Graph Metrics.}
\label{sec:method:evaluation metrics-lg}

Computed values in large graph requires Monte Carlo sampling, therefore evaluation metrics also need to be adapted accordingly for output/edge/node.
Within $n$ random draws in computation space $\mathbb{Z}$, $i_{th}$ draw $z^{(i)}$ is associated with output $o^{(i)}$, edge $e^{(i)}$, and nodes $u^{(i)}$ and $v^{(i)}$.
For output $k$'s privacy, we have the estimation $\hat{m_o(k)} = \frac{1}{|\{i : o^{(i)}=k\}|} \sum_{i : o^{(i)}=k} f(z^{(i)})$. 
Similarly, for edge $l$ and node $v$ privacy, we have $\hat{m_e(l)} = \frac{1}{|\{i : e^{(i)}=l\}|} \sum_{i : e^{(i)}=l} f(z^{(i)})$ and $\hat{m_n(v)} = \frac{1}{|\{i : u^{(i)}=v \lor v^{(i)}=v\}|} \sum_{\substack{i : u^{(i)}=v \lor v^{(i)}=v}} f(z^{(i)})$.  
The estimation for $\overline{m_o}$ is expressed as $\hat{\overline{m_o}} = \frac{1}{|O_n|} \sum_{k \in O_n} \hat{m_o(k)}$, $|O_n|$ denotes the number of distinct outputs randomly selected from the output set $O$ within $n$ draws.
Similarly we have the estimation $\hat{\overline{m_e}} = \frac{1}{|E_n|} \sum_{l \in E_n} \hat{m_e(l)}$ for edges, and $\hat{\overline{m_n}} = \frac{1}{|V_n|} \sum_{v \in V_n} \hat{m_n(v)}$ for nodes.
Since in the computation space, each output/edge/node is equally likely to be selected associated with random samples, the linear relation exists between $T$ and  $\hat{\overline{m_o}}$, 
which is $\hat{\overline{m_o}} = \frac{1}{|O_n| \times |\{i : o^{(i)}=k\}|} $$\left( \sum_{k \in O_n}  \sum_{i : o^{(i)}=k} f(z^{(i)}) \right) 
= \frac{T}{|O_n| \times |\{i : o^{(i)}=k\}|}$.
Similarly for $\hat{\overline{m_e}}$ and $\hat{\overline{m_n}}$.
For $\overline{\epsilon_o}$, we don't have this strict linear relation.

\section{Experiments \& Results}
\label{sec:experiment-and-evalustion}

\subsection{Setup}
\label{sec:setup}

\subsubsection{\textbf{Datasets (G)}}
Datasets used in our experiments are listed in Table~\ref{tab:dataset}. In Karate Club~\cite{zachary1977information}, nodes are members and edges are friendships. HIV~\cite{auerbach1984cluster} represents individuals and their sexual contacts. CEO~\cite{faust1997centrality} includes individuals/organizations as nodes and their relationships as edges. In EU-Email~\cite{yin2017local}, nodes represent people and edges represent email communications.
We also report graph statistics including vertex count, edge count, density and normalized rank (= $rank(A)/|V|$).
In discussing performance, we will sometimes refer generically to ``large graphs'', by which we mean datasets with $\ge 1000$ nodes.
\begin{table}[!t] 
    \caption{Dataset Statistics}
    \begin{center}
    \begin{tabular}{c|c|c|c|c}
        \toprule
        \textbf{Dataset} & \textbf{$|$V$|$} & \textbf{$|$E$|$} & \textbf{Density} & \textbf{Nor. Rank} \\ 
        \toprule
        Karate Club              & 34 & 78 & 0.14 & 0.71\\ 
        HIV              & 40 & 41 & 0.05 & 0.80 \\ 
        CEO             & 40 & 95 & 0.12 & 0.75\\
        EU-Email             & 1005 & 16706 & 0.03 & 0.98\\ 
        \bottomrule
    \end{tabular}
    \end{center}
    \label{tab:dataset}
\end{table}

\subsubsection{\textbf{Graph Synthesizers (M)}}
\label{sec:synthesizer}
We consider two different model families, statistical graph synthesizers (SGS) and deep-learning based graph synthesizers (DGS).
We select several statistical graph synthesizers, including the spectral graph forge (SGF)~\cite{baldesi2018spectral}, Chung-Lu model (CL)~\cite{chung2002connected}, block two-level Erd\H{o}s–R\'enyi model (BTER)~\cite{kolda2014scalable} and stochastic block model (SBM)~\cite{karrer2011stochastic} for the following reasons.
First, these models are widely used in diverse domains. SBM is applied in recommendation systems~\cite{godoy2016accurate} and community detection, CL is used to model brain networks~\cite{agarwala2023geometric}, BTER supports anomaly detection~\cite{bridges2015multi} and biological network modeling, and graph spectra plays a crucial role in data compression~\cite{loukas2019graph} and graph approximation.
Second, these models serve as important building blocks for private graph synthesizers. For example, PrivGraph~\cite{yuan2023privgraph} uses CL for graph reconstruction, LDPGen~\cite{qin2017generating} leverages BTER for synthetic graph generation, and GraphWorld~\cite{palowitch2022graphworld} employs a variant of SBM to generate diverse datasets. SGF~\cite{baldesi2018spectral} and GNS~\cite{zeng2021gns} utilize graph spectra to enhance data anonymity. 
For deep graph synthesizers, we consider graph autoencoders (GAE)~\cite{kipf2016variational}, variational graph autoencoders (VGAE)~\cite{kipf2016variational} and adversarially regularized graph autoencoders (ARGA)~\cite{pan2018adversarially}, which are widely used in different applications. 

\subsubsection{\textbf{Hyper-parameters}}
\label{sec:hyper-params}
Before calculating the $\delta$-bound for \plrv{} values, we determine a suitable $\delta$ value. In this work, we set $\delta = \frac{1}{|V|}$ for the following reasons.
Firstly, to measure the worst case, $\delta$ should be sufficiently small so that the majority of \plrv{}s fall below the $\delta$-bound. 
Setting $\delta$ in this way not only helps to exclude outliers but also effectively captures the worst case.
Secondly, as graph size increases, individuals become less distinguishable within the dataset. In other words, more nodes may share the same degree, making it easier to blend in and be less unique. So a smaller $\delta$ imposes a stricter privacy guarantee to maintain the balance. 
Finally, setting $\delta$ to zero is impractical, as it would make privacy quantification too sensitive to extreme/outlier values.

\begin{figure}[!t]
\centering
\includegraphics[width=0.8\linewidth, alt={boxplots display sample size validation across iterations, showing convergence and stability of privacy and utility metric as iterations increases.}]{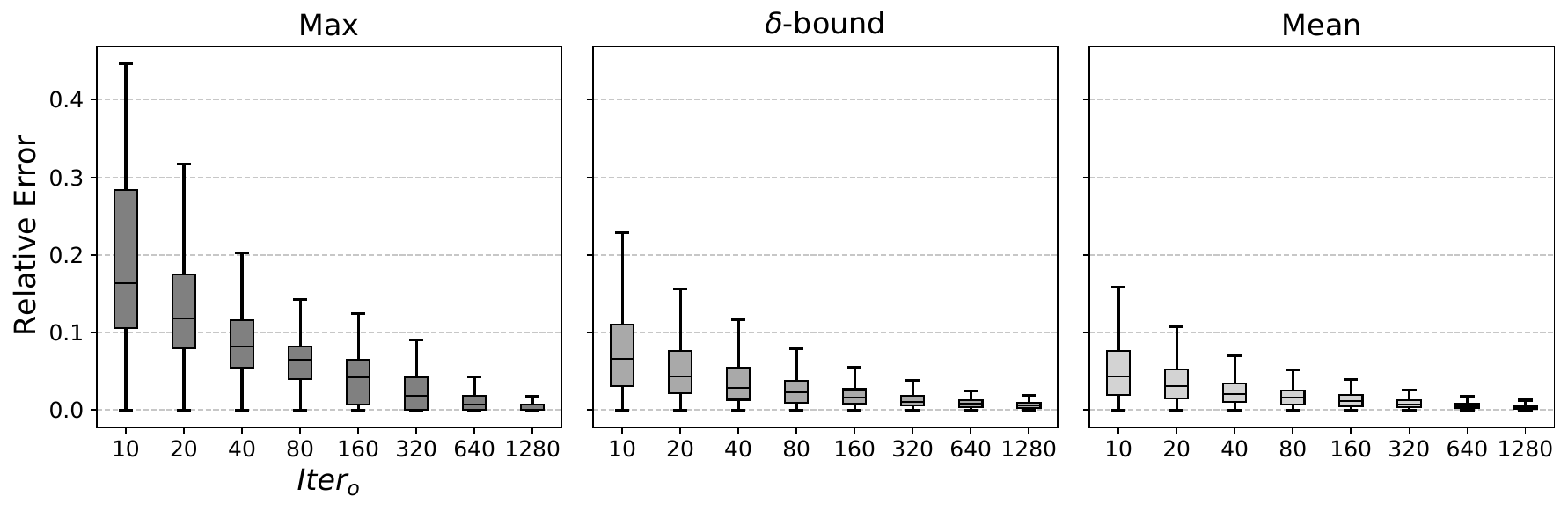}
\caption{\textbf{Sample Size Validation ($Iter_o$).}}
\label{fig:validation-numo}
\end{figure}

The evaluation is based on multiple outputs from $P_A$. Given that data curators may have limited computational resources, we conducted an experiment to identify a suitable output iteration ($Iter_o$). 
We use setup ($G$:Karate Club $\times$ $M$:BTER) for testing, since evaluation results of BTER exhibit the most randomness compared with other $M$ (BTER scatter plots in Figure~\ref{fig:scatter-all} show more dispersion). 
For validation, we varied $Iter_o$ exponentially and tested 1000 random trials for each $Iter_o$. We calculated the relative error using the max \plrv{}, as it captures the full range of \plrv{}s, also other values include $\delta$-bound and mean. 
As shown in Figure~\ref{fig:validation-numo}, relative error converges to 0 when $Iter_o$ exceeds 320 and before 640. Therefore, we set $Iter_o=500$ for subsequent experiments to balance precision and computational cost.

\begin{figure*}[!t] 
\centering
\includegraphics[width=\linewidth, alt={Comparison chart illustrating the performance of exact solutions versus Monte Carlo approximation methods with increasing iterations, showing the scatter group shift.}]{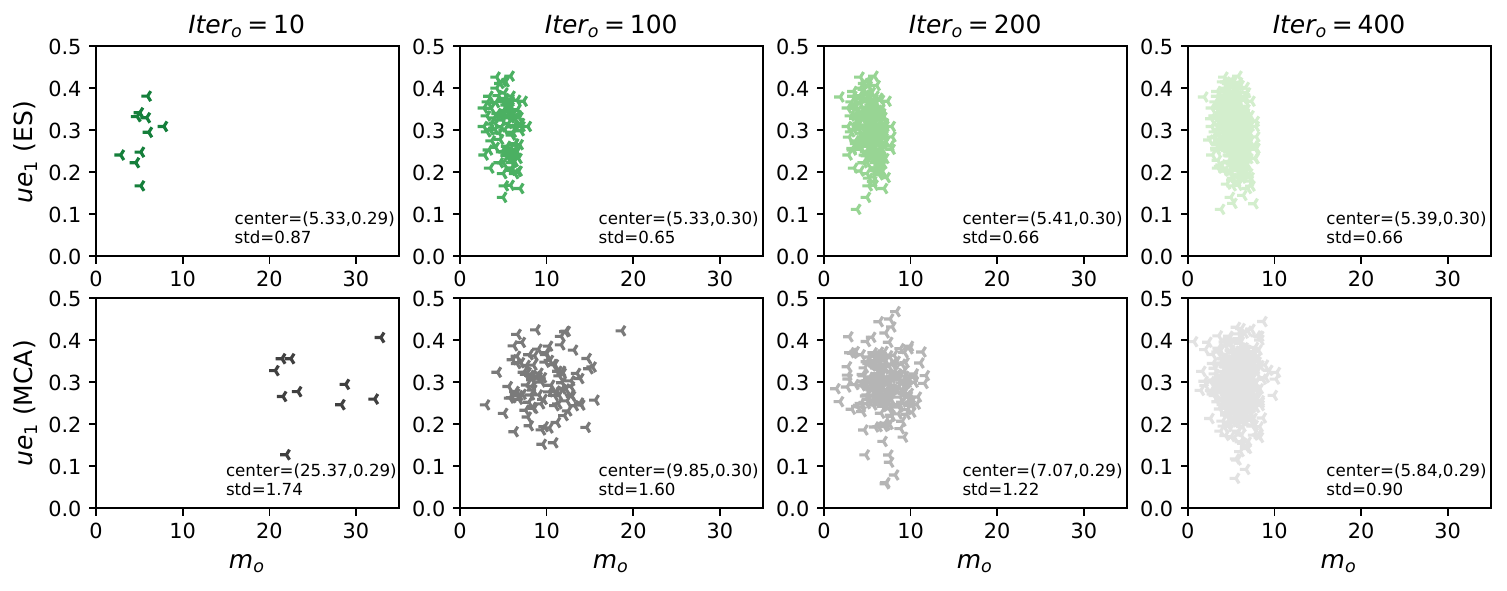}
\caption{\textbf{Exact Solution vs. Monte Carlo Approximation.}}
\label{fig:estimated-pa}
\end{figure*}

\subsection{Results \& Findings}
Next, we show  evaluation results across datasets and graph synthesizers, which span in two model families statistical graph synthesizers (SGS) and deep graph synthesizers (DGS). Sections~\ref{sec:small-graph-scenarios}-\ref{sec:small-graph-en} evaluate SGS on small graphs, Section~\ref{sec:small-graph-DGS} evaluates DGS on small graphs, Section~\ref{sec:res:large-graph} evaluates SGS/DGS on large graph dataset.

\subsubsection{\textbf{Exact Solution vs. Monte Carlo Approximation}}
\label{sec:small-graph-scenarios}
ES is with full access to $P_A$, while MCA is based on the estimated version $\hat{P}_A$ derived from multiple observed outputs.
We select setup ($G$:Karate Club $\times$ $M$:BTER) for its high variability of output evaluation results (see Figure~\ref{fig:scatter-all}, BTER scatter plots are more dispersed than other $M$), and focus on $ue_1$ as modularity effectively distinguishes $M$.
Figure~\ref{fig:estimated-pa} presents evaluation results for different $Iter_o$ under ES/MCA scenarios. 
The upper figures show that in the ES scenario, scatter plots within a group stabilize in a fixed area as $Iter_o$ increases and get denser (center stabilizes and std decreases then converges). 
We expect that as $Iter_o \to \infty$, there will be more scattering around this fixed area.
The lower figures show the MCA scenario. Points within a group move to the left as $Iter_o$ increases (center moves to the left and std decreases).
Notably, as $Iter_o$ increases, MCA evaluation results converge to ES. For MCA with large $Iter_o$ (e.g., 400), MCA evaluation results closely match those of ES. 
This implies that when the evaluation framework sets a proper $Iter_o$ (large enough), MCA evaluation results are close to ES, which takes less computation but greater model access (less safe).
Further experiments are based on the ES scenario (with $Iter_o=500$, validated in Section~\ref{sec:hyper-params}), which takes less computation and offers more accurate evaluation.

\textbf{Key Takeaways.}  
MCA evaluation results converge to ES when we set a proper/large $Iter_o$.  
MCA requires more computation but less model access, making it safer to model, while ES demands less computation but more access. 
In practice, ES can provide evaluation results faster and precisely, we continue the experiments in ES.

\subsubsection{\textbf{Within/Across Statistical Graph Synthesizer (SGS)}}
\label{sec: results: acrossm}
Table~\ref{tab:vector-all} (left, within one SGS)  summarizes the vector evaluation results for one setup ($M$:SGF $\times$ $G$:Karate Club) with varying $\alpha$ (model parameter). 
Output privacy loss metrics $(\overline{\epsilon_o},\overline{m_o})$ and $\epsilon_{mn}$ increase sharply between $\alpha = 0.5$ and $0.7$, then keep stable. 
At the same $\alpha$ value, utility error metric vector drops to zero, indicating a significant structural shift in synthetic graphs. Notably, this transition point ($\alpha=0.7$) aligns with $G$ (Karate Club)'s normalized rank of 0.7.
Additionally, as $\alpha$ increases, average-case privacy metrics $(\overline{m_e}, \overline{m_n})$ show a slight increase after transition point 0.7, while worst-case metrics $(\overline{\epsilon_e}, \overline{\epsilon_n})$ decreases. This suggests that with larger $\alpha$, the privacy level for edges and nodes becomes worse on average, which aligns with theoretical expectation.

\begin{table*}[!t]
    \caption{Setup Evaluation Vectors within/across $M$ ($G$: Karate Club $\times$ $M$: Statistical / Deep Graph Synthesizers).}
    \centering
    \small
    \setlength{\tabcolsep}{4pt} 
    \begin{tabular}{c | *{5}{w{c}{0.8cm}} | *{4}{w{c}{0.8cm}} | *{3}{w{c}{0.8cm}}}
        \toprule
        & \multicolumn{9}{c|}{Statistical Graph Synthesizers (SGS)} & \multicolumn{3}{c}{Deep Graph Synthesizers (DGS)} \\
        \cmidrule(lr){2-10} \cmidrule(lr){11-13}
        & \multicolumn{5}{c|}{Within one SGS (SGF, with different $\alpha$)} & \multicolumn{4}{c|}{Across different SGS} & \multicolumn{3}{c}{Across different DGS} \\
        \cmidrule(lr){2-6} \cmidrule(lr){7-10} \cmidrule(lr){11-13}
        V\textbackslash $M$ & 0.1 & 0.3 & 0.5 & 0.7 & 0.9 & CL & BTER & SBM & SGF & GAE & VGAE & ARGA \\
        \midrule
        $\overline{m_{o}}$        & 1.51  & 2.22 & 2.15 & 3.61 & 4.23 & \underline{0.32} & 5.48  & 5.15  & 2.15 & 2.52 & 0.0  & 10.98 \\
        $\overline{\epsilon_{o}}$  & 6.67  & 7.90 & 7.00 & 23.03 & 23.03 & \underline{1.71} & 32.24 & 31.37 & 7.00 & 6.86 & 0.03 & 21.14 \\
        $\overline{m_{e}}$        & 2.78  & 2.35 & 2.13 & 2.80 & 3.25 & \underline{0.32} & 4.19  & 3.62  & 2.13 & 2.26 & 0.0  & 10.30 \\
        $\overline{\epsilon_{e}}$  & 7.62  & 6.23 & 5.67 & 2.80 & 3.25 & \underline{1.95} & 9.39  & 8.47  & 5.67 & 5.91 & 0.03 & 18.40 \\
        $\overline{m_{n}}$        & 2.92  & 2.42 & 2.39 & 2.61 & 3.26 & \underline{0.36} & 3.82  & 3.27  & 2.39 & 2.35 & 0.0  & 10.36 \\
        $\overline{\epsilon_{n}}$  & 10.18 & 7.77 & 7.31 & 7.24 & 8.80 & \underline{2.18} & 17.96 & 16.6  & 7.31 & 7.14 & 0.03 & 20.10 \\
        $\epsilon_{mn}$            & 6.75  & 8.06 & 7.17 & 23.03 & 23.03 & \underline{1.75} & 30.28 & 29.25 & 7.17 & 7.83 & 0.04 & 23.13 \\
        \midrule
        $\overline{ue_1}$ & 0.25 & 0.11 & 0.08 & \underline{0.01} & \underline{0.01} & 0.29 & 0.30 & \underline{0.08} & \underline{0.08} & 0.67 & 0.74 & 0.61 \\
        $\overline{ue_2}$ & 0.39 & 0.18 & 0.11 & \underline{0.} & \underline{0.} & 0.53 & 0.50 & 0.56             & \underline{0.11} & \underline{0.13} & \underline{0.12} & \underline{0.11} \\
        $\overline{ue_3}$ & 0.43 & 0.23 & 0.17 & \underline{0.} & \underline{0.} & 0.23 & 0.29 & 0.28             & \underline{0.17} & 6.04 & 8.06 & 5.46 \\
        $\overline{ue_4}$ & 0.08 & 0.04 & 0.02 & \underline{0.} & \underline{0.} & 0.11 & 0.26 & 0.94             & \underline{0.02} & 1.01 & 0.98 & 0.93 \\
        $\overline{ue_5}$ & 0.16 & 0.07 & 0.03 & \underline{0.} & \underline{0.} & 0.19 & 0.29 & 0.78             & \underline{0.03} & 0.92 & 0.98 & 0.80 \\
        $\overline{ue_6}$ & 0.31 & 0.13 & 0.08 & \underline{0.} & \underline{0.} & 0.44 & 0.41 & 0.78             & \underline{0.08} & 0.93 & 0.99 & 0.83 \\
        \bottomrule
    \end{tabular}
    \label{tab:vector-all}
\end{table*}

\begin{figure*}[!t] 
\centering
\includegraphics[width=\linewidth, alt={Scatter plot matrix showing privacy utility evaluation metrics per output across synthesizers. The top row displays average-case privacy and the bottom row displays worst-case privacy, label-independent utility on the left and label-dependent utility on the right.}]{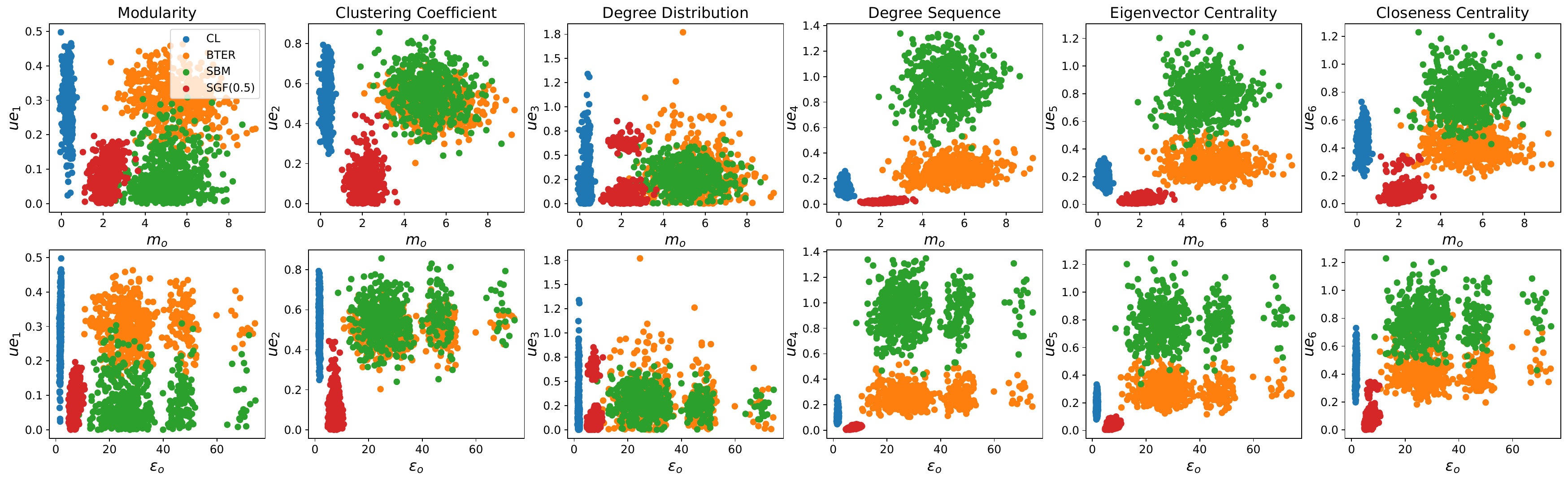}
\caption{\textbf{Evaluation per Output across $M$.}  
(top: average-case privacy, bottom: worst-case privacy; left: label-independent utility, right: label-dependent utility.)}
\label{fig:scatter-all}
\end{figure*}

Table~\ref{tab:vector-all} (middle, across different SGS) summarizes vector evaluation results across $M$ ($M$:CL/BTER/SBM/SGF(0.5) $\times$ $G$:Karate Club). 
$\epsilon_{mn}$ varies widely across the mechanisms (range around 30, from CL to BTER), but much less within a single mechanism (range around 15 in SGF with $\alpha$ varying), indicating that privacy loss is primarily influenced by the choice of mechanism.
For output/edge/node privacy evaluation, all privacy loss metrics follow the order CL$<$SGF$<$BTER/SBM, which reflects different mechanisms' design. CL preserves local properties (degree distribution), results in lower privacy loss. In contrast, SGF and BTER/SBM capture more detailed community structure, leading to higher privacy loss. 
Specifically, SGF approximates community information through low-rank approximation, while BTER/SBM employ community detection algorithms (e.g. Louvain Method) that optimize modularity over the entire graph, resulting in the highest privacy loss among tested mechanisms.
For label independent utility metrics $\overline{ue_1}$-$\overline{ue_3}$, SBM/SGF yield the lowest modularity errors, as both are designed to preserve community structure with modularity as a key metric. SGF achieves lowest error for the clustering coefficient, reflecting its effectiveness in capturing tightly knit groups. CL/SGF perform best in preserving degree distribution, since CL is tailored to preserve degree distribution and SGF closely approximates the adjacency matrix.     
For label dependent utility metrics $\overline{ue_4}$-$\overline{ue_6}$, SGF/CL outperform BTER/SBM, as SGF/CL better preserve node dependent features. While BTER/SBM better preserve block-level structures, these are less aligned with localized, label-dependent features.

\textbf{Unfold per Output.}
Figure~\ref{fig:scatter-all} shows evaluation results across $M$, it unfolds more details for each output in different privacy-utility evaluation space. 
BTER/SBM scatter plots exhibit greater dispersion indicating less robustness, while CL/SGF produce more stable evaluation results across output. 
Moreover, CL/SGF point clouds are closer to the axis, showing better performance than BTER/SBM in privacy-utility evaluation. 
Additionally, though $\overline{ue_3}$ for BTER/SBM appears close (0.29/0.28) in setup vectors (in Table~\ref{tab:vector-all}, middle, $\overline{ue_3}$), scatter plot (in Figure~\ref{fig:scatter-all}, $ue_3$ column) reveals BTER exhibits greater randomness than SBM. This emphasizes that beyond average performance, stability of outputs should guide the choice of $M$.

\begin{figure}[!t]
\centering
\includegraphics[width=\linewidth, alt={Two-row heatmap matrix displaying risk fraction metrics RF3 in the top row and RF4 in the bottom row across different synthesizer models M, highlighting privacy risk variation across different privacy utility metric threshold configurations.}]{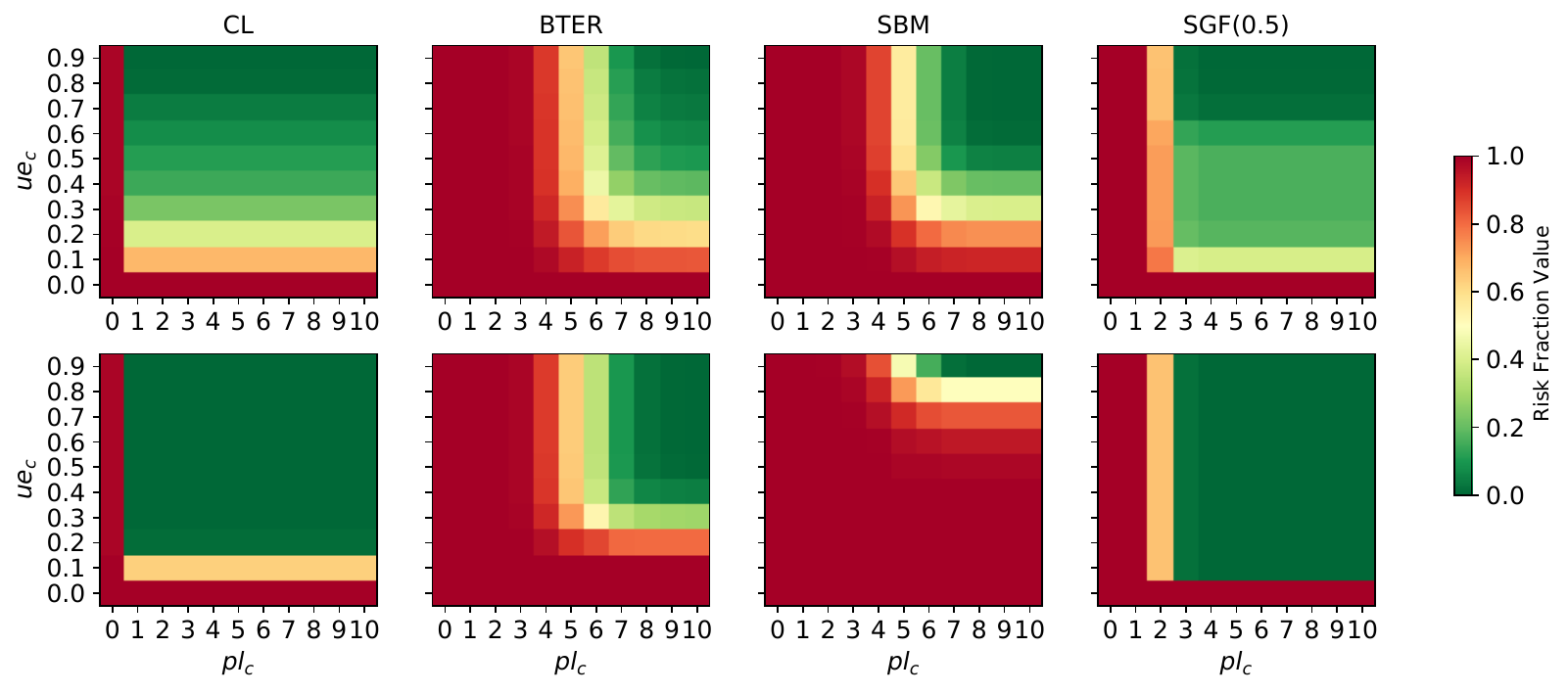}
\caption{\textbf{Risk Fraction Heatmap across $M$.} (1st row: $RF_3$, 2nd row: $RF_4$)} 
\label{fig:rf-34}
\end{figure}

Next, we unfold more details in degree distribution and degree sequence with average-case privacy evaluation, and dive deeper into difference between label dependent and label independent utility metrics. 
Figure~\ref{fig:rf-34} presents the risk fraction in utility $(ue_3, ue_4)$ and privacy $(m_o)$ under varying constant constraints, with red indicating higher risk. 
We set $pl_c$ interval as $[0,10]$ and $ue_c$ as $[0,1]$, since these intervals effectively differentiate the evaluation results across $M$.    
Several key observations exist. First, CL/SGF generally outperform BTER/SBM, as indicated by the larger green areas. Second, CL/SGF produce more stable results, evidenced by fewer yellow cells between red/green. 
Third, although CL does not achieve lowest average utility errors ($\overline{ue_3}, \overline{ue_4}$) in vector form Table~\ref{tab:vector-all}, it attains the lowest risk (with most green cells in heatmaps) compared with other $M$. 
This finding indicates under same risk constraints, CL outperforms other $M$, 
it also emphasizes the importance of fine-grained empirical evaluation with intuitive visualization.  
Additionally, though BTER/SBM have similar risk results for degree distribution ($RF_3$), BTER outperforms SBM in degree sequence ($RF_4$) by more green cells. It shows BTER is better than SBM at preserving label dependent utility metrics, this advantage is also reflected in $\overline{ue_{4}}$-$\overline{ue_6}$ in Table~\ref{tab:vector-all} (middle).

\textbf{Key Takeaways.}
For performance among four mechanisms, SGF/CL achieves better performance. SGF has both moderate privacy loss and lowest utility error, and CL offers the lowest privacy loss and moderate utility error. SBM yields highest utility error and high privacy loss, and BTER exhibits both the highest privacy loss and high utility error. For output robustness, SGF/CL are more reliable than BTER/SBM. 
Last, scatter plots are effective for evaluating output individually. Risk heatmaps provide insights about privacy-utility trade-off from risk angle with fixed constraints, by analyzing the behavior of grouped outputs.

\subsubsection{\textbf{Edge \& Node Privacy}}
\label{sec:small-graph-en}

\begin{figure*}[!t] 
\centering
\includegraphics[width=\linewidth, alt={Paired boxplot compares edge and node privacy loss across datasets G and synthesizer M. The left plot groups privacy loss by dataset G, while the right plot groups edge privacy loss by synthesizer M, across different datasets.}]{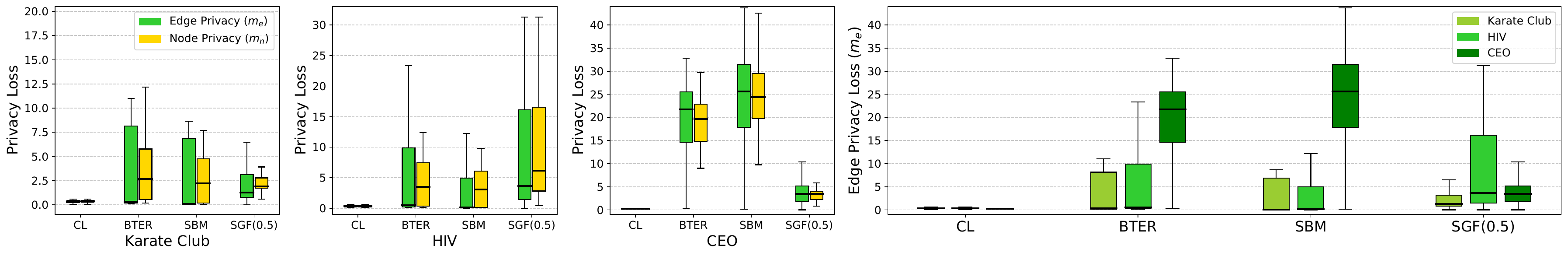}
\caption{\textbf{Edge/Node Privacy Loss across $G$ and $M$.} (left: grouped by $G$; right: grouped by $M$, edge privacy.)}
\label{fig:enp-all}
\end{figure*}

Figure~\ref{fig:enp-all} shows edge and node privacy loss across mechanisms and datasets. 
In Figure~\ref{fig:enp-all} (left), node privacy loss is consistently less variable than edge privacy loss (e.g., for BTER on HIV dataset, $m_e$ ranges around 24, while $m_n$ is around 12). 
Figure~\ref{fig:enp-all} (right) summarizes edge privacy loss grouped by $M$. 
Across $M$, BTER/SBM have the highest privacy loss, SGF is intermediate and CL is lowest. 
This ranking reflects mechanism design: BTER/SBM use community detection and preserve detailed structure, increasing the risk of revealing edge location. In contrast, CL preserves expected node degrees, a global property that reduces edge privacy loss, and SGF approximates the adjacency matrix.

We note two interesting observations. First, privacy loss order of $M$ varies across $G$. For Karate Club/CEO dataset, the order is CL$<$SGF$<$BTER/SBM, while for HIV dataset, it is CL$<$BTER/SBM$<$SGF (see Figure~\ref{fig:enp-all}, left).
Although these datasets have similar numbers of nodes, their densities vary a lot. Denser graphs tend to contain more complex structures, allowing community detection methods to capture more details and resulting in higher privacy loss. This accounts for the increased privacy loss of BTER/SBM on denser graphs like Karate Club/CEO.
Another interesting finding is in BTER/SBM, privacy loss increases with graph size (see Figure~\ref{fig:enp-all}, right, Karate Club with lowest $m_e$ and CEO with highest).
It aligns with our expectation since more nodes/edges enrich graph structures. 

Furthermore, we examine the hypothesized association between high privacy loss entities and high-degree nodes, using recall to quantify how well this relationship holds. 
For each $G$, setups with the highest privacy loss are Karate Club$\times$BTER, HIV$\times$SGF, and CEO$\times$SBM, as shown in Figure~\ref{fig:enp-all} (left). We define high-privacy-loss entities using threshold $m_e$/$m_n$ for edges/nodes and identify high-degree nodes with median degree.
Recall is computed by $\frac{|V_{Deg-H} \cap V_{Pri-H}|}{|V_{Priv-H}|}$ for nodes and $\frac{|E_{Deg-H} \cap E_{Pri-H}|}{|E_{Pri-H}|}$ for edges ($V_{Deg-H}$ represents high-degree nodes, $E_{Deg-H}$ represents edges with at least one node in high-degree nodes group). 
For three setups, edge/node recall results are [1.0, 0.63], [1.0, 0.21], and [0.63, 0.60], indicating that high privacy loss entities are strongly associated with high-degree nodes (with most recalls exceeding 0.5).

\textbf{Key Takeaways.}
Privacy loss tends to be concentrated among nodes but more dispersed across edges. 
Typical order of edge privacy loss across mechanisms is BTER/SBM $>$SGF$>$CL. However, this order may vary slightly across datasets, depending on graph statistics (e.g., graph size, density, and normalized rank).
Furthermore, high-degree nodes are often associated with entities that exhibit high privacy loss.

\subsubsection{\textbf{Deep Graph Synthesizer (DGS)}}
\label{sec:small-graph-DGS}

We extend the evaluation to deep learning-based graph synthesizers. For the model configuration, we adapt the original configuration to the 16-dim hidden layer and 8-dim embedding layer, learning rate 0.01 and epochs 200 keep the same. We summarize the results into Table~\ref{tab:vector-all} (right).
For the privacy evaluation, ARGA shows the highest privacy loss and VGAE shows the least privacy loss, GAE is in the middle. 
This is because GAE is a trainable graph encoder with graph reconstruction loss as optimization goal, it can leak the information of original graph. 
ARGA adds a standard discriminator on top of the encoder, in order to fix the overfitting flaw in traditional GAE. 
As a result, it increases privacy loss through the training process and less utility error compared with GAE. 
VGAE adds KL-divergence penalty on top of GAE, which forces the node vector to be close to a normal distribution. As a result, the output graphs are more indistinguishable and cause less privacy loss. 
The privacy evaluation results align with the graph synthesizer design.
For the utility evaluation, VGAE shows the most utility error compared with GAE/ARGA, GAE performs worse than ARGA. 

Furthermore, our evaluation framework also unifies the comparison between statistical graph synthesizer and deep ones.  
Comparing the evaluation results from two model families, deep graph synthesizers can lead to greater privacy loss than statistical families. 
This is because the current DGS configuration overwhelms the data size (small), causing more privacy loss for the input graphs.
For utility performance, SGS generally outperforms DGS, except on $\overline{ue_2}$ clustering coefficient metric. This advantage likely stems from deep models’ ability to capture latent triangle via message passing, a capability that most statistical models lack.

\textbf{Takeaways.}
Deep graph synthesizers exhibit varying privacy levels due to inherent differences in model design, and DGS can lead to higher privacy loss compared with SGS in small graphs.
DGS generally underperform SGS across most utility metrics, except for the clustering coefficient (see Table~\ref{tab:vector-all}, right).

\subsubsection{\textbf{Large Graphs}}
\label{sec:res:large-graph}
Recall that to apply our framework to large graphs, we modified Algorithm \ref{alg:quantifyeneis} to utilize Monte Carlo sampling in computation space. In this section, first, we validate the sampling approach, and then we apply it to large graph dataset.

\begin{figure}[htbp] 
  \centering
  \includegraphics[width=\linewidth, alt={Grouped figures show sampling convergence and scalability results for different datasets. Columns a and b display computation convergence in relative error across datasets G and synthesizer models M as sample size grows. Row c shows the convergence of estimated m bar o on the left, and relative error scaling as a function of graph node size N on the right.}]{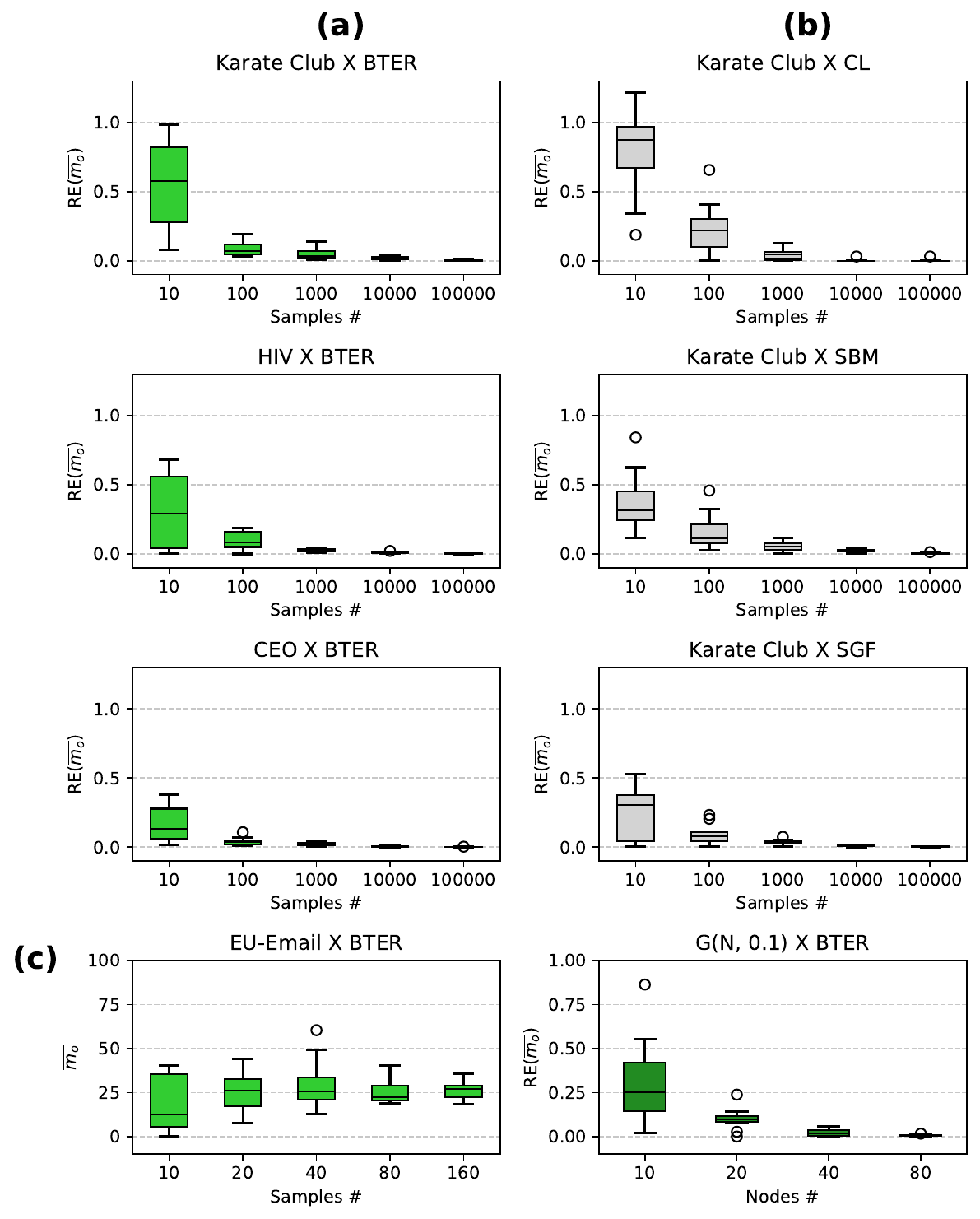}
  \caption{\textbf{Sampling Results.} Column (a) and (b) show the computation convergence in relative error across $G$ and $M$. Row (c) left shows the convergence of estimated $\overline{m_o}$; right shows the relative error changing with graph size $N$.    
  }
  \label{fig:large-graph-res}
\end{figure}

\textbf{Validation of Sampling Approach.} To validate the accuracy of this sampling approach, we monitor the random samples by the relative error of $\hat{\overline{m_o}}$ in computation space. 
Figure~\ref{fig:large-graph-res} Column (a) and (b) detail how the relative error changes with increasing sample size, across datasets and mechanisms. For each sample size we conducted 10 parallel tests; results shown by box plots.
In this experiment, the sample size was increased exponentially, while the $Iter_o$ was held constant at 500. 
Relative errors fall quickly with sample size, while precision increases.  For instance, with a sample size = 10 in setup Karate Club $\times$ BTER, the relative error ranges from 0.1 to 1.0, declining by a factor of approximately 10 with a sample size of 1000.
Column (a) shows that across different datasets, a sample size as small as 100 is sufficient to estimate $\overline{m_o}$ with a relative error below 0.2.
Column (b) illustrates that the relative errors for all tested mechanisms converge towards zero, and stabilize once the sample size exceeds 100. 
Convergence suggests that sample sizes of 100-1000 draws are sufficient for achieving highly accurate estimation, not only across different datasets but also for various mechanisms. 

\begin{figure}[!t] 
  \centering
  \includegraphics[width=0.5\linewidth, alt={Boxplot displays testing results across different iterations, plotting metric performance and convergence trends as the iteration increases.}]{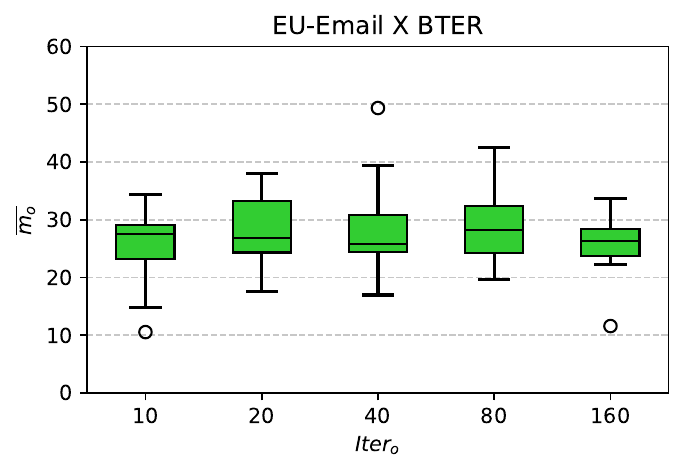}
  \caption{\textbf{$Iter_o$ Testing.}}
  \label{fig:iter-o-testing}
\end{figure}

We then extended the evaluation to a large dataset EU-Email, where obtaining the ground truth of $\overline{m_o}$ was computationally infeasible due to resource limitations. 
Drawing upon our previous findings on convergence, we set the maximum sample size for this experiment to 160 (larger than 100). 
Figure~\ref{fig:iter-o-testing} provides the stability analysis and shows that the estimated privacy metric $\overline{m_o}$ remains stable for a sample size of 160, even with exponentially increasing $Iter_o$. Therefore, a value of $Iter_o$=80 was chosen for our experiment. 
Consequently, we estimated $\overline{m_o}$ directly from uniform sampling. The results are presented in Figure~\ref{fig:large-graph-res} Row (c) left. 
As the sample size increases, we observe that the interval of the estimated values narrows, which serves as strong evidence that the computation is converging toward a stable value within a computationally accessible sample size. 
The privacy metrics reported below were computed using a stabilized sample size (160), $Iter_o$ (80) and the final reported values are the average of 10 tests.

\textbf{SGS vs. DGS.}
We summarize the results of EU-Email in Table~\ref{tab:sd-largegraph}, covering privacy utility evaluation on statistical and deep graph synthesizers, with average metric and uncertainty score.
In large graph experiments, DGS provide much better privacy than SGS, different results are shown in small graph (see Table~\ref{tab:vector-all}, where DGS can reach to higher privacy loss). 
Additionally, the privacy loss difference between SGS and DGS are pronounced in large graph results, SGS can reach to much higher privacy loss (around 50) than DGS (close to 0). 
This is because the current DGS configuration (8-dim embedding layer) is bottleneck, which forces DGS to drop detailed structure.
As a result, utility error of DGS is much higher than SGS.
Besides, for uncertainty measurement of privacy evaluation metrics, DGS shows higher scores than SGS. 
This is because DGS learn graph structure through stochastic optimization and parameter space, which increase the variability in generated structures and make the privacy metric less stable.

\begin{table*}[!t] 
    \caption{Setup Evaluation Vectors within/across $M$. ($G$: EU-Email $\times$ $M$: Statistical Graph Synthesizer/ Deep Graph Synthesizer), with average metric and relative uncertainty score.}
    \centering
    \begin{tabular}
    {c|cccc|ccc}
        \toprule
        \multicolumn{5}{c|}{Statistical Graph Synthesizer} & \multicolumn{3}{c}{Deep Graph Synthesizer} \\
        \midrule
        V\textbackslash $M$ & CL & BTER & SBM & SGF &  GAE & VGAE & ARGA \\ 
        \midrule
        $\overline{m_{o}}$ & 0.13(0.13) & 24.59(0.08) & 47.18(0.06) & 33.13(0.03) & 0.(0.) & 0.01(0.51) & 0.33(0.23) \\
        $\overline{\epsilon_{o}}$  & 0.38(0.05) & 49.86(0.07) & 99.57(0.05) & 50.91(0.03)  & 0.(0.) & 0.05(0.08) & 0.80(0.22) \\
        $\overline{m_{e}}$  & 0.13(0.09) & 25.12(0.07) & 47.99(0.05) & 32.85(0.02) & 0.(0.) & 0.(0.52) & 0.36(0.22) \\
        $\overline{\epsilon_{e}}$  & 0.13(0.09) & 25.12(0.07) & 47.99(0.05) & 32.85(0.02) & 0.(0.) & 0.(0.52) & 0.36(0.22) \\
        $\overline{m_{n}}$  & 0.13(0.09) & 24.84(0.06) & 47.63(0.05) & 32.49(0.02) &  0.(0.) & 0.(0.59) & 0.37(0.22) \\
        $\overline{\epsilon_{n}}$   & 0.17(0.08) & 27.42(0.07) & 54.14(0.05) & 35.18(0.02)  & 0.(0.) & 0.01(0.23) & 0.42(0.22) \\
        $\epsilon_{mn}$   & 1.59(0.05) & 326.03(0.14) & 629.72(0.02) & 135.34(0.02) & 0.(0.) & 0.43(0.13) & 30.68(0.21) \\
        \midrule
        $\overline{ue_1}$  & 0.71 & 0.32 & \underline{0.01} & 0.15  &  0.94 & 0.94 & 0.93 \\
        $\overline{ue_2}$   & 0.63 & 0.72 & 0.84 & 0.43 &  0.23 & 0.23 & 0.37 \\
        $\overline{ue_3}$   &\underline{0.05} & 0.53 & 1.23 & 0.21 &  18.91 & 18.91 & 18.67 \\
        $\overline{ue_4}$  &0.01 & 0.04 & 0.93 & \underline{0.} &  1. & 0.92 & 0.47 \\
        $\overline{ue_5}$  &0.05 & 0.10 & 0.86  & \underline{0.}  &  1. & 0.93 & 0.51 \\
        $\overline{ue_6}$  & 0.28 & 0.10 & 0.86 & \underline{0.03} & 1. & 0.93 & 0.52\\
        \bottomrule
    \end{tabular}
    \label{tab:sd-largegraph}
\end{table*}

\textbf{SGS Privacy/Utility.}
For utility, each SGS model excels at preserving specific graph property: SBM best maintains modularity due to its community detection step, CL preserves degree distribution by maintaining individual node connection probabilities, and SGF is effective at preserving label dependent features. 
Compared with small graphs results ($\overline{ue}$ rows in Table~\ref{tab:vector-all}, left), utility preservation is more pronounced in large graph ($\overline{ue}$ rows in Table~\ref{tab:sd-largegraph}, left). For instance, $\overline{ue_1}$ in SBM drops from 0.08 in small graphs to 0.01 in large graphs, and $\overline{ue_3}$ in CL decreases from 0.23 to 0.05, and $\overline{ue_4}$ in SGF drops from 0.02 to 0.
For privacy loss metrics (in Table~\ref{tab:sd-largegraph}, left), output/edge/node privacy results follow the same ranking, CL performs best with the lowest privacy loss, SBM highest, and BTER/SGF fall in between. 
However, the order is different from some small graph results. For EU-Email/Karate Club/CEO datasets, SBM shows more privacy loss than SGF, while in the HIV dataset, SGF shows more privacy loss than SBM (see Figure~\ref{fig:enp-all}). 
SGF shows more privacy loss in graphs with lower density and low normalized rank (e.g., HIV); while SBM shows more privacy loss in graphs with high normalized rank or high density. For example, Table~\ref{tab:dataset} shows EU-Email's normalized rank is 0.98 which is much higher that other tested datasets, and Karate Club or CEO have higher density that others (above 0.1, others much less than 0.1). 
This difference arises because in sparse and structurally simple graphs, eigenvalues/eigenvectors used in SGF react sharply to edge perturbations, leading to more privacy loss. while in denser or more complex graphs, perturbing the block can influence many edges' generation, which causes higher privacy loss.

In summary, for large graphs, DGS provide better privacy and worse utility compared with SGS. 
Within SGS, CL model provides better privacy which is comparable to DGS, CL also provides better utility compared with DGS.
Among SGS, SBM provides the worst utility and the worst privacy, while SGF provides the best utility and moderate privacy. CL is with moderate utility and the best privacy, BTER with moderate utility and privacy. 
Data curators can refer to our results for model selection with similar graph statistics.

\textbf{Graph Size.}
Furthermore, we investigated the influence of graph size on the privacy evaluation. For this experiment, we utilized graphs generated by Erd\H{o}s–R\'enyi model (ER). 
The ER model is particularly suitable as it allows for precise control over graph properties. To test the effect of scale, we held the graph density constant at 0.1 while varying the number of nodes.
Figure~\ref{fig:large-graph-res} Row (c) right shows that the relative error of $\overline{m_o}$ consistently decreases as the number of nodes increases.
The trend shows the inherent stability of our privacy evaluation framework when applied to the larger graphs.
The explanation for this phenomenon lies in the relationship between graph size and the definition of edge-based privacy loss. With a fixed density, a larger graph contains a quadratically greater number of edges. Privacy loss is defined by the effect of adding/removing a single edge.
In larger graph, the structural contribution of any single edge becomes less significant to the total set.
This "dilution effect" means the privacy loss values associated with individual edge modification, become more uniform across edges within the larger graph. 
As a result, the estimation from random samples is less variable, which leads to a lower relative error.
This finding serves as a guideline for the user, the framework yields increasingly stable evaluation results as graph size increases, providing greater confidence in the evaluation metric estimation.

\textbf{Key Takeaways.} 
(i) Our framework provides a computationally efficient method for evaluating privacy of mechanisms on large graphs. By leveraging the uniform random sampling, it makes direct comparison between mechanisms practical.
(ii) 
Our framework's reliability in evaluating large graphs is supported by two distinct observation. 
First, as the number of samples increases, the estimated privacy metrics stabilize to a constant. Second, this stability is independently reinforced by the “dilution effect,” a property of large graphs where the privacy impact of any single edge is diminished. 
These two trends indicate that a reliable approximation can be reached. 
(iii) The framework can be used to generate specific, evidence-based recommendations. For instance, in the large, sparse graph, we recommend the CL model for data curators who prioritize privacy and the SGF model for those who require higher data utility.
(iv) The relative size of between DGS model capacity and data size influences the privacy loss comparison between SGS/DGS.
(v) DGS exhibits more uncertainty due to its stochastic optimization in learning process, compared with SGS.

\section{Related Work}
\label{sec:related-work}

\textbf{Graph Data \& Differential Privacy.}
Differential privacy is a formal framework to protect individual privacy in data publishing~\cite{dwork2006differential}. In graph data type, DP can be achieved in different levels~\cite{wang2013preserving}: edge differential privacy protects the presence of edges; similar for node-DP.
Many methods have been proposed to achieve DP in various tasks (sensitive queries) on graphs (e.g. degree distribution~\cite{karwa2012differentially}, subgraph counting~\cite{chen2013recursive}, edge weight~\cite{li2017differential}).
However, previous methods focus on a specific query instead of a generic query (synthetic graph like ours). 
Graph synthesizers aim to generate synthetic graphs that preserve key properties of real-world graphs. 
E.g., Pygmalion~\cite{sala2011sharing} uses a dK-graph model to preserve the number of connected k-node subgraphs. LDPGen~\cite{qin2017generating} explores decentralized setting in graphs and preserves community structures. PrivGraph~\cite{yuan2023privgraph} generates synthetic graphs by exploiting original community information. 
PGS extend specific queries to generic synthetic graph query, but they do not evaluate privacy for the synthesis process before adding DP noise. Instead, PGS integrate directly data synthesizers and DP noise~\cite{liu2024pgb}).
Our framework setting is based on graph data and synthetic graph queries, which is more generic. Moreover, we focus on empirical evaluation based on actual outputs. 

\textbf{Auditing Privacy of Synthetic Data Generation.}
There are different privacy auditing tools for different data types.  
For tabular data, \cite{annamalai2024you} audits privacy by membership inference attacks (to determine whether a specific individual's data was included to generate synthetic data). 
\cite{houssiau2022framework} uses math framework to test whether synthetic data reveals anything beyond what is allowed by generator card. 
\cite{giomi2024unified} translates the requirements in privacy protection regulations to executable attack algorithm to quantify the privacy leakage. 
For text data, \cite{panda2024privacy} audits privacy for LLM by member inference attack and crafted canary. 
For graph data, structural re-identification~\cite{narayanan2008robust} de-anonymizes synthetic graphs using structural pattern; link inference predicts hidden relationships between nodes~\cite{zheleva2009join}; membership inference identifies if a node/edge was in the original graph~\cite{olatunji2021membership}.    
Privacy loss random variables connect actual outputs with measurable privacy loss. Earlier studies~\cite{doroshenko2022connect, dandekar2020differential} have applied \plrv{} to calculate privacy loss on tabular data and counting queries. Inspired by these works, we extend use of privacy loss random variable and approximate differential privacy to graph data, allowing for privacy evaluation with synthetic graph output.

\textbf{Graph Models \& Structural Utility.}
Extensive research focuses on developing graph generative models designed to produce synthetic graphs that are statistically indistinguishable from their empirical counterparts.
These models prioritize structural similarity, accurately replicating key properties such as degree distributions.
The model categories range from classical probabilistic approaches to more recent, deep learning techniques. 
In the probabilistic category, the Chung-Lu model~\cite{chung2002connected} preserves the expected degree sequence, stochastic block model~\cite{karrer2011stochastic} maintains the community structure of the original graph, and stochastic kronecker graph model~\cite{leskovec2007graph} replicates real graph patterns through recursive structures. 
\cite{10.1007/978-3-319-11257-2_12} use exponential family random graph models to generate networks with similar labeled and unlabeled properties to a target.
Modern methodologies employ graph neural network within frameworks like variational autoencoders~\cite{kipf2016variational, simonovsky2018graphvae} and generative adversarial networks~\cite{pan2018adversarially, bojchevski2018netgan} to learn and replicate the graph patterns.
While utility evaluation is crucial, this focus often leaves privacy implications of the generation process unexamined, creating the need for rigorous privacy evaluation frameworks.

Our work differs from the above work in two aspects. 
First, unlike studies that focus on specific graph statistics such as degree distribution or subgraph counts, ours is designed for a more general setting, evaluating privacy of the entire graph generation process. 
Second, our approach is empirical, analyzing the privacy loss from actual output graphs. This evaluation is formally grounded in the definitions of privacy loss random variable and approximate differential privacy. 
Third, our evaluation framework can be applied in a broad class of graph synthesizers, both statistical and deep-learning based families.

\section{Discussion \& Limitations}
\label{sec:discussion/limitation}
Our framework evaluates a graph model where edges are generated independently, functioning as a Bernoulli graph synthesizer. Although we do not consider all classes of synthesizers, Bernoulli class is an extremely important case for privacy analysis. 
In particular, models with edgewise dependencies will, \emph{ceteris paribus}, exhibit greater privacy loss, as dependencies by definition create mutual information between edges variables. 
This creates additional routes for an attacker to infer structural patterns from partially observed data, versus the case in which edges are independently drawn.
It thus follows that independence models are of particular interest for privacy-preserving graph synthesis, as they (\emph{ceteris paribus}) minimize deductive disclosure risk from partial observations. 
Likewise, independence models are typically more scalable than models with dependence, making them more relevant to production of proxy graphs from large data sets. These features and the inherent simplicity, make this generators class a natural target for analysis.

A potential limitation of our work is the high cost of exact computation of privacy evaluations for large graphs. 
We achieve scalability, however, by the use of Monte Carlo sampling.  While the resulting estimates are not exact, they are computationally consistent (i.e., can be made to approach arbitrarily close to the ground truth in the limit of computational effort) and unbiased. 
As we have shown, high precision can be obtained for reasonable sample sizes even for large graphs, making this analysis practical beyond the small-graph regime. 
Besides, partial graph observation and dataset continuous uploaded setting are beyond our scope.

\textbf{Deployment \& Guidelines.}
We recommend that the framework be operated by authorized data curators or researchers, who not only can access the private dataset, but also has the appropriate resources to run the evaluations. 
Additionally, evaluations should be conducted in controlled computing environments, such as secured institutional servers or research clusters.
Users of the framework should ensure that input graph data complies with applicable privacy regulations, ethical guidelines or data-use agreements before running evaluations. 

Different users have different objectives. Data curators need the privacy evaluation of individual output before they release the synthetic graph, or for model selection. Data contributors are primarily concerned with the privacy of their social relationships and can use the evaluation results to understand potential disclosure risks. Researchers can leverage the framework to study the privacy-utility tradeoff, compare graph synthesizers, and guide better synthesizer design. 
As a general guideline, users with sufficient computational resources should apply the framework directly to their datasets and candidate graph synthesizers, as privacy and utility characteristics may vary across datasets and model configurations. This enables a customized assessment tailored to the intended data release scenario.
When direct evaluation is infeasible, users may instead refer to the key findings reported in our evaluation section. By comparing the structural characteristics of their datasets with those evaluated in this work, they can identify graph synthesizers with privacy and utility properties that best align with their release goal.

\section{Conclusion} 
\label{sec:conclusion}
In this paper, we propose \framework{} to enable different granular privacy/utility evaluations, based on the actual outputs. 
Specifically for privacy, we leverage privacy loss random variables to build the computational space and approximate differential privacy to extract key evaluation metrics. 
Extensive experiments (on four real-world datasets and seven widely-used graph synthesizers) show how to evaluate setups with our framework from different granularity levels, including per output, per entity and the whole setup. 
Next, we analyze the evaluation results by linking graph statistics and synthesizers' design, providing insight for further private graph synthesizer selection/design.   
To the best of our knowledge, this is the first effort to connect graph synthesizer to empirical evaluation, with the formal privacy definition.
Furthermore, our framework is extendable to other graph queries (e.g. path length) or data types (e.g. location data).

\section*{Acknowledgement}
Yu Duan, Luca Baldesi and Athina Markopoulou were supported by NSF awards 1956393 and 1900654, and UCI SSoE Samueli Graduate Fellowship.

\section*{Ethics and Privacy Statement}
We used publicly available graph datasets, which do not raise ethical issues. Our work focuses on privacy evaluation of graph synthesizers, which contributes to measure or even enhance the current privacy level of graph models and datasets.
This work provides actionable insights to stakeholders. Evaluation results from our framework help data curators make responsible model choices, balancing privacy and utility. By offering tailored evaluations for researchers and data contributors, the framework promotes transparency and supports the development of safer, trustworthy data systems.

\bibliographystyle{unsrt}  
\bibliography{references} 

\end{document}